\documentclass[letterpaper,floatfix,aps,prb,amsmath,amssymb,twocolumn,superscriptaddress,nopacs]{revtex4-2}
\usepackage{graphicx}
\usepackage{verbatim}
\usepackage{color}
\usepackage{array}
\usepackage{placeins}
\usepackage{cancel}
\usepackage[pdfauthor={Fischer, Catelani},
            pdftitle={Excess quasiparticles and their dynamics in the presence of subgap states}]{hyperref}

\hypersetup{pdfstartview={XYZ null null 1.05}}

\renewcommand{\Im}{\mathrm{Im\,}}
\renewcommand{\Re}{\mathrm{Re\,}}
\renewcommand{\vec}[1]{\mathrm{\mathbf{#1}}}

\usepackage{diagbox}
\begin{document}
\title{Excess quasiparticles and their dynamics in the presence of subgap states}

\author{P. B. Fischer}

\affiliation{JARA Institute for Quantum Information (PGI-11), Forschungszentrum J\"ulich, 52425 J\"ulich, Germany}

\affiliation{JARA Institute for Quantum Information, RWTH Aachen University, 52056 Aachen, Germany}

\author{G. Catelani}

\affiliation{JARA Institute for Quantum Information (PGI-11), Forschungszentrum J\"ulich, 52425 J\"ulich, Germany}

\affiliation{Quantum Research Center, Technology Innovation Institute, Abu Dhabi 9639, UAE}

\date{\today}

\begin{abstract}
    Material inhomogeneities in a superconductor generically lead to broadening of the density of states and to subgap states. The latter are associated with spatial fluctuations of the gap in which quasiparticles can be trapped. Recombination between such localized quasiparticles is hindered by their spatial separation and hence their density could be higher than expectations based on the recombination between mobile quasiparticles. We show here that the recombination between localized and mobile excitations can be efficient at limiting the quasiparticle density. We comment on the significance of our findings for devices such as superconducting resonators and qubits. We find that for typical aluminum devices, the subgap states do not significantly influence the quasiparticle density.
\end{abstract}

\maketitle

A finite energy gap for excitations is what makes superconductors attractive materials for realizing electronic devices with low losses. At temperature $T$ approaching the critical one ($T_c$), quasiparticle excitations are a significant source of losses, but because of the gap, at $T\ll T_c$ the number of quasiparticles is exponentially suppressed in thermal equilibrium. However, experimental evidence points to a low-temperature density of quasiparticles much larger than expected in devices such as qubits and resonators. In fact, various non-equilibrium mechanisms are being investigated as sources of excess quasiparticles, such as environmental and cosmic radiation~\cite{Vepsalainen.2020,Cardani.2021,Wilen.2021}, pair-breaking photons~\cite{Houzet.2019,Diamond.2022,Pan.2022,Liu.2024}, and phonon bursts~\cite{RAP.2024,Yelton.2025}.
The basic model for the dynamics of the quasiparticle density $n$ as function of time $t$ was introduced long ago by Rothwarf and Taylor (RT)~\cite{Rothwarf.1967}: quasiparticles created at a rate $G$ recombine pairwise, leading to a rate equation $dn/dt=G-R n^2$ with $R$ the recombination coefficient; in the steady-state $dn/dt=0$, the density is determined by the competition between generation and recombination, $n=\sqrt{G/R}$.  
More recently~\cite{Bespalov.2016,Grunhaupt.2018} it has been argued that a thorough understanding of the experimental evidence requires extending this simple model, in particular to account for inhomogeneities in the superconducting gap, for which there is direct evidence in strongly disordered superconductors~\cite{Sacepe.2008}; in fact the devices studied in Ref.~\cite{Grunhaupt.2018} were fabricated with highly disordered granular aluminum. The gap inhomogeneities can be due to spatial variations in the concentration of magnetic impurities~\cite{Kulik.1968} or in the strength of the pairing constant~\cite{Larkin.1972}. 
For concreteness, in this work we focus on weak magnetic impurities, but the results are straightforwardly applicable to the other case.

\begin{figure}[!b]
    \centering
    \includegraphics[width=\columnwidth]{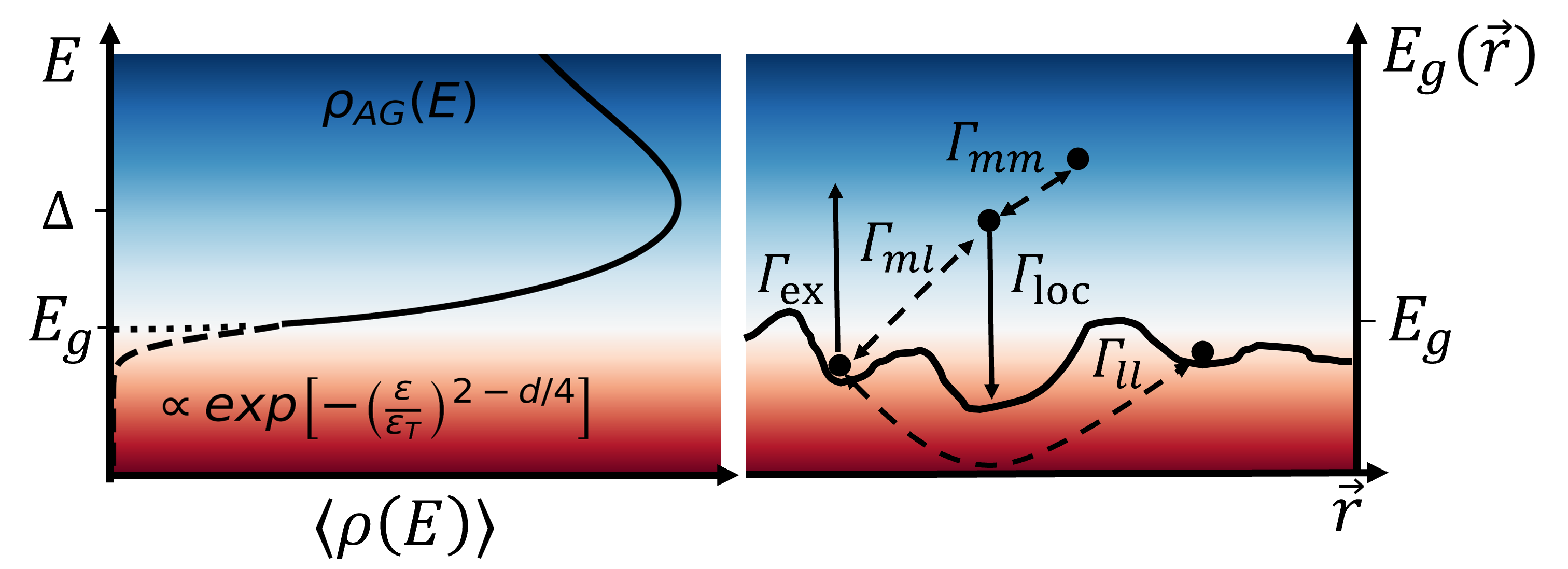}
    \caption{Left: Average density of states as function of energy, showing the Abrikosov-Gorkov broadened peak above $E_g$ and the subgap tail below it.  Right: Local value of the gap as function of position; also depicted are the various processes affecting the quasiparticle densities, namely recombination between localized (red background, recombination coefficient $\Gamma_{ll}$) and/or mobile (blue background, $\Gamma_{ml}$ and $\Gamma_{mm}$) quasiparticles, localization (rate $\Gamma_\mathrm{loc}$), and excitation ($\Gamma_\mathrm{ex}$).}
    \label{fig: rates}
\end{figure}

Magnetic impurities have long been know to suppress $T_c$ with increasing concentration~\cite{Matthias.1958}; here we are interested in how they affect the density of states (DoS). According to BCS theory, in the absence of magnetic impurities the DoS is characterized by a square-root divergent peak at an energy $\Delta$. Using the approach of Abrikosov and Gorkov (AG)~\cite{AG.1961,Skalski.1964}, extended to account for fluctuations in the concentration of impurities~\cite{Silva.2005,Skvortsov.2013,Fominov.2016}, it can be shown that a sufficiently high concetrnation of magnetic impurities weakly coupled to the conduction electrons~\cite{Fominov.2016} modifies the DoS in two ways (see Fig.~\ref{fig: rates} left): first, they broaden the peak by changing the square root divergence to a square root threshold at energy $E_g = \Delta (1-\eta^{2/3})^{3/2} < \Delta$; this broadening depends on the average impurity concentration and is quantified by a dimensionless pair-breaking parameter $\eta=1/\tau_s \Delta$, where $1/\tau_s$ is related to the exchange interaction part of the total scattering rate of electrons by impurities; we set the reduced Planck constant $\hbar$ and the Boltzmann constant $k_B$ to one throughout. Second, spatial fluctuations in the impurity concentration cause local depressions of the energy gap where quasiparticles can be trapped (see Fig.~\ref{fig: rates} right), since the states in these local depression are subgap (that is, they have energy smaller than $E_g$); the subgap states add to the DoS a compressed-exponential ``tail'' decaying over an energy scale $\epsilon_T\ll \eta^{2/3}\Delta$. According to Ref.~\cite{Bespalov.2016}, the spatial separation between ``localized'' quasiparticles trapped in the subgap sates hinders their recombination; this reduction of the quasiparticle recombination rate then leads to an increase in their number above what is expected for ``mobile'' quasiparticles of energy $E>E_g$. In this work we revisit the role of localized quasiparticles in determining the overall quasiparticle density and discuss the implications of our results for superconducting devices.

A phenomenological model generalizing the RT one can be straightforwardly written down for the dynamics of the densities of localized ($x_l$) and mobile ($x_m$) quasiparticles~\cite{Grunhaupt.2018,supplemental}
\begin{align}\label{eq: Effective Rate Equation}
    \frac{d x_{l}}{d t} & = \Gamma_\mathrm{loc} x_m  - \Gamma_\mathrm{ex} x_l - \Gamma_{ml} x_m x_l - \Gamma_{ll} x_l^2 + g_l \\ 
    \frac{d x_{m}}{d t} & = \Gamma_\mathrm{ex} x_l  - \Gamma_\mathrm{loc} x_m - \Gamma_{ml} x_m x_l - \Gamma_{mm} x_m^2 + g_m \label{eq:xm rate eq}
\end{align}
Here, the densities $x_\alpha=n_\alpha/n_\mathrm{Cp}$, $\alpha = l,\, m$, are normalized by that of the Cooper pairs $n_\mathrm{Cp}=2\nu\Delta$, with $\nu$ the DoS per spin at the Fermi energy, $\Gamma_\mathrm{ex}$ denotes the rate for excitation from a localized to a mobile state, $\Gamma_\mathrm{loc}$ the rate for the inverse (localization) process, $\Gamma_{\alpha\beta}$, $\alpha,\, \beta = l,\, m$, the recombination coefficient between localized ($l$) and/or mobile ($m$) quasiparticles (see Fig.~\ref{fig: rates} right), and $g_{\alpha}$ the rate at which quasiparticles are generated by some pair-breaking mechanism. In the absence of localized states ($x_l = 0 = \Gamma_\mathrm{loc}$), the model reduces to the RT equation. In contrast, the authors of Ref.~\cite{Bespalov.2016} assumed that quasiparticles are generated at high energy only ($g_l = 0$) and quickly localize ($\Gamma_\mathrm{loc} \gg \Gamma_{ml} x_l,\,\Gamma_{mm} x_m$); with these assumptions, and if excitation can be ignored, Eqs.~\eqref{eq: Effective Rate Equation} and \eqref{eq:xm rate eq} approximately reduce to $dx_l/ dt = g_m - \Gamma_{ll} x_l^2$, again taking the RT form. Although direct comparison with the approach of that work is not straightforward, its main insight can be qualitatively expressed by saying that effectively the recombination rate $\Gamma_{ll}$ depends on the generation rate through the dimensionless parameter $\kappa = (n_\mathrm{Cp} r_c^3)^2 g_m/\Gamma_{mm}$, where $r_c$ 
denotes the relevant radius of the localized states, which
is expected to be of order the coherence length $\xi$ for the disorder strengths at which, as we will show, this regime can be relevant and a few times $\xi$ for weaker disorder. 
Then for $\kappa \gg \kappa_c$ Ref.~\cite{Bespalov.2016} concludes that $\Gamma_{ll} \simeq \Gamma_{mm}$, while for $\kappa \ll \kappa_c$ the relationship becomes $\Gamma_{ll} \simeq \Gamma_{mm} g(\kappa)$ with the function $g(\kappa)$, whose concrete form in not relevant for our purposes, being always smaller than unity. 
The cross-over value $\kappa_c$ was estimated from simulations of the recombination process between localized quasiparticles to be of order $\kappa_c \simeq 10^{-4}$. The arguments in Ref.~\cite{Bespalov.2016} were developed using formulas for bulk superconductors which can be generalized to effectively two-dimensional films of thickness less than the coherence length~\cite{supplemental}.We note that strictly speaking $\Gamma_{ll}$ accounts for recombination between quasiparticles located in different traps; recombination within a trap could be enhanced (cf. Ref.~\cite{deRooij.2024}), but this does not affect our arguments~\cite{supplemental}.

In considering the relevance of the results of Ref.~\cite{Bespalov.2016} to experiments, one should examine whether the assumption of fast localization is justified, since the relaxation rate of a quasiparticle due to phonon emission decreases strongly with energy upon approaching the superconducting gap~\cite{Kaplan.1976}. More recently, it has been shown that the absorption of low-energy photons can lead to a finite width (that is, an ``effective temperature'') of the quasiparticle distribution even if the phonons are assumed to be at zero temperature~\cite{G.Catelani.2019}; similarly, high-energy photons responsible for quasiparticle generation generally lead to a distribution with finite effective temperature~\cite{Fischer.2024}.
A natural question then is under which conditions relating the (effective) temperature and the typical energy $\epsilon_T$ of the subgap states can localization significantly affect quasiparticle dynamics. The goal of this paper is to study such dynamics taking into account states both above and below $E_g$. To this end, we need microscopic estimates for the coefficients entering Eqs.~\eqref{eq: Effective Rate Equation} and \eqref{eq:xm rate eq}; such estimates can be obtained using a kinetic equation approach (see for instance Ref.~\cite{Larkin.1986}). Here we discuss the values of the rates, presenting more details in the Supplemental Material~\cite{supplemental}.

The recombination of two quasiparticles is accompanied by the emission of a phonon. The strength of the electron-phonon interaction is typically quantified by a time $\tau_0$~\cite{Kaplan.1976,G.Catelani.2019,Fischer.2024} in terms of which we have for the recombination coefficient of mobile quasiparticles $\Gamma_{mm}\simeq r$ with $r=4(\Delta/T_c)^3/\tau_0$. In thin superconducting films the parameter $r$ can be enhanced due to so-called phonon trapping~\cite{Chang.1978}; it could also be affected by the concentration of implanted impurities~\cite{Barends.2009}. Therefore, we will assume that $r$ is determined experimentally for a given material and film thickness. 
Moreover, using the approach of Ref.~\cite{Bespalov.2016} we find that $\Gamma_{ml} \simeq \Gamma_{mm}$. Regarding $\Gamma_{ll}$, as discussed above we have $\Gamma_{ll} \le \Gamma_{mm}$; this inequality is also compatible with the possible ineffectiveness of phonon-trapping enhancement of recombination for localized quasiparticles~\cite{deRooij.2024}. Clearly, a finite value for $\Gamma_{ll}$ can only lead to a lower density of localized quasiparticles, so an upper bound for $x_l$ is obtained by setting $\Gamma_{ll} = 0$, while a lower bound by letting $\Gamma_{ll}=\Gamma_{mm}$. Concerning generation, similarly to Ref.~\cite{Bespalov.2016} we set $g_l=0$; we
discuss in the Supplemental Material~\cite{supplemental},
for both thermal phonons and pair-breaking photons, under which conditions  $g_l \ll g_m$, so 
that our assumption gives
a reasonable approximation. 

Turning to the localization and excitation rates, we note that the latter vanishes in the limit of zero phonon temperature $T$ and assuming that no nonequilibrium mechanism such as stray photons can give energy to the trapped quasiparticles. Below we will consider first the case $\Gamma_\mathrm{ex}=0$ 
and then the effect of a finite excitation rate, in particular due to thermal phonons. The localization rate can be estimated from~\cite{supplemental}
\begin{equation}\label{eq:Glocdef}
    \Gamma_\mathrm{loc}x_m \simeq \frac{2}{\Delta }\int\limits_{E_g}^\infty \!dE \, \rho_{AG}(E)f(E)\tau^{-1}_\mathrm{loc}(E)
\end{equation}
where $E$ is the energy measured from the Fermi level, $\rho_{AG}$ is the AG DoS (cf. Fig.~\ref{fig: rates} left), and $f$ is the quasiparticle distribution function; the energy-dependent localization time $\tau_\mathrm{loc} \propto \tau_0$ accounts for the electron-phonon interaction and depends on the (disorder-averaged) density of states of the localized states (normalized by the normal-state DoS $\nu$)~\cite{Larkin.1972,Silva.2005,Skvortsov.2013,Fominov.2016,supplemental} 
\begin{equation}\label{eq:rhol}
    \rho_l(\tilde{\epsilon}) \simeq \frac{a_d}{\eta^{2/3}}  \left(\frac{\epsilon_T}{\Delta}\right)^{1/2} \tilde{\epsilon}^{\alpha_d} \exp[-\tilde{\epsilon}^{2-d/4}]
\end{equation}
with $d=2,\,3$ the effective dimensionality of the system ($d=2$ for films of thickness less than the coherence length), $\alpha_d=[d(10-d)-12]/8$, $a_2 \simeq 0.32$, $a_3 \simeq 0.53$, and $\tilde{\epsilon} = (E_g-E)/\epsilon_T$, where the energy scale $\epsilon_T$ depends on the strength of the disorder and its fluctuations~\cite{supplemental}. For quasiparticles with energy close to $E_g$ ($E_g< E \lesssim \Delta$) we estimate
\begin{align}\label{eq:tloc_final}
    \tau_\mathrm{loc}^{-1} (\epsilon) &\approx  r\frac{a_d}{8-d}\left(\frac{\epsilon_T}{\Delta}\right)^{7/2} \left[\mathit{\Gamma}\left(\frac{4(\alpha_d+1)}{8-d}\right) \epsilon^2 \right. \\ & \left. +2 \mathit{\Gamma}\left(\frac{4(\alpha_d+2)}{8-d}\right) \epsilon + \mathit{\Gamma}\left(\frac{4(\alpha_d+3)}{8-d}\right) \right] \nonumber
\end{align}
where $\epsilon = (E-E_g)/\epsilon_T$ and the symbols $\mathit{\Gamma}$ within square brackets denote the gamma function. A more general discussion of the relation between localization rate and electron-phonon interaction is given in the Supplemental Material~\cite{supplemental}.

If the time $\tau_\mathrm{loc}$ were independent of energy, we would simply have $\Gamma_\mathrm{loc} = 1/\tau_\mathrm{loc}$. Here for our purposes we establish an upper bound on $\Gamma_\mathrm{loc}$ by considering the competition between localization and relaxation of a mobile quasiparticle into mobile states. Considering again quasiparticles with energy close to $E_g$, the rate $\tau_m^{-1}$ for the latter process is
\begin{equation}
    \tau_m^{-1}(\epsilon) \simeq r \frac{4}{105}\sqrt{\frac{2}{3}} \left(\frac{\epsilon_T}{\Delta}\right)^{7/2} \epsilon^{7/2}
\end{equation}
We define the crossover (normalized) energy $\epsilon_c$ by the equation $\tau_\mathrm{loc}(\epsilon_c) = \tau_m(\epsilon_c)$ [$\epsilon_c\simeq 3.55\, (2.32)$ for $d=3\,(2)$]; quasiparticles with energy $\epsilon>\epsilon_c$ will more likely remain mobile after emitting a phonon rather than localize, while the opposite holds for $\epsilon < \epsilon_c$. 
Therefore, the relevant energy range determining localization can be taken between $\epsilon =0 $ ($E=E_g$) and $\epsilon\sim\epsilon_c$.
Since $\tau_\mathrm{loc}^{-1}$ is an increasing function of energy, we conclude that $\Gamma_\mathrm{loc} \lesssim \tau_\mathrm{loc}^{-1}(\epsilon_c)$; in what follows, we will take $\Gamma_\mathrm{loc}$ to be given by the upper bound, 
\begin{equation}\label{eq:gammaloc}
\Gamma_\mathrm{loc}\simeq b_d r (\epsilon_T/\Delta)^{7/2},
\end{equation}
with $b_3 \simeq 2.62$ and $b_2 \simeq 0.59$.

Next, we study the steady-state density as predicted by Eqs.~\eqref{eq: Effective Rate Equation} and \eqref{eq:xm rate eq} when $\Gamma_\mathrm{ex}=0$. Let us assume that the generation rate is large enough that $\Gamma_{ll} \simeq r$; then taking the sum of the two equation we find the steady-state total density $x=x_m+x_l =\sqrt{g_m/r}$, independent of the localization rate. However, the densities of the two components depend on the latter,
\begin{equation}\label{eq:xmxlbeta}
    x_m = \sqrt{\frac{g_m}{r}} \frac{1}{1+\beta}, \quad x_l=\sqrt{\frac{g_m}{r}} \frac{\beta}{1+\beta}
\end{equation}
with $\beta = \Gamma_\mathrm{loc}/\sqrt{g_m r}$. When $\beta \ll 1$, most quasiparticles are mobile and the density of localized quasiparticles is small, $x_l \simeq \Gamma_\mathrm{loc}/r \ll x_m$. In fact, in the regime of small $\beta$ the effect of $\Gamma_{ll}$ can be ignored (at leading order); this can be understood by noticing that the recombination rate for mobile quasiparticles $\tau_r^{-1} = r x_m \simeq \sqrt{g_m r}$ is large compared to $\Gamma_\mathrm{loc}$, and therefore most quasiparticles recombine before they have a chance to localize. 

\begin{figure}[bt]
    \centering
    \includegraphics[width=\columnwidth]{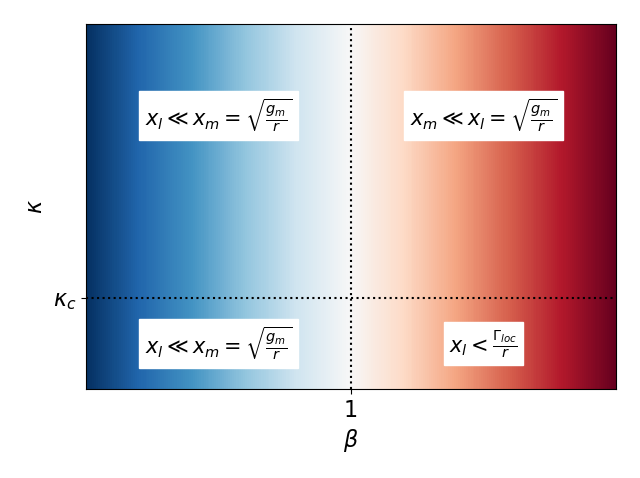}
    \caption{Diagram representing the various regime possible depending on the values of parameters $\beta$ and $\kappa$. The background color denotes quasiparticles being mostly mobile (blue, $\beta < 1$) or localized (red, $\beta > 1$). 
    Only when $\beta>1$ and $\kappa<\kappa_c$ (lower right quadrant), the density can be influenced by the suppression of recombination between localized quasiparticles discussed in Ref.~\cite{Bespalov.2016}.
    }
    \label{fig: pd}
\end{figure}

As the generation rate decreases, the regime $\beta \gtrsim 1$ can be reached. 
The expressions in Eq.~\eqref{eq:xmxlbeta} still apply so long as $\kappa \gtrsim \kappa_c$;
interestingly, when both $\beta$ and $\kappa$ are large, we get $x_m \simeq g_m/\Gamma_\mathrm{loc}$ and the localized states effectively acts as quasiparticle traps, which would beneficial for qubits~\cite{Riwar.2016}.
If $\kappa \lesssim \kappa_c$ and $\beta \gtrsim 1$ the 
mechanism discussed in Ref.~\cite{Bespalov.2016} could become effective at suppressing $\Gamma_{ll}$ below $r$; 
since in this fast localization regime we expect $x_m<x_l$ and the density to decrease monotonically with decreasing generation rate, an upper limit on $x_l$ when $\kappa \lesssim \kappa_c$ and $\beta \gtrsim 1$ is always given by its value estimated when these parameters are of order unity, namely
\begin{equation}\label{eq:xl_ub}
    x_l \lesssim \Gamma_\mathrm{loc}/r
\end{equation}
We stress that only if the (total) density is below this upper limit, the mechanism considered in Ref.~\cite{Bespalov.2016} could be relevant, since if the density is higher, $x_l$ is determined by the competition between localization of mobile quasiparticles and localized-mobile recombination, while localized-localized recombination can be ignored. We summarize the possible regime for the quasiparticle densities in Fig.~\ref{fig: pd}.

We now consider the effect of a mechanism exciting quasiparticles from localized to mobile states. If $\kappa \gg \kappa_c$, the results in Eq.~\eqref{eq:xmxlbeta} generalize to
\begin{equation}\label{eq:xmxlbetatilde}
    x_m = \sqrt{\frac{g_m}{r}} \frac{1+\tilde{\beta}}{1+\beta+\tilde{\beta}}, \quad x_l=\sqrt{\frac{g_m}{r}} \frac{\beta}{1+\beta+\tilde{\beta}}
\end{equation}
where $\tilde{\beta}=\Gamma_\mathrm{ex}/\sqrt{g_m r}$. Not surprisingly, the excitation process increases $x_m$ at the expense of $x_l$. 
Concerning the value of the excitation rate $\Gamma_\mathrm{ex}$, we note that
at sufficiently low temperatures, $T \ll \epsilon_T$, phonon absorption is not effective at delocalizing quasiparticles; then absorption of photons with energy $\omega_0 \gg \epsilon_T$ should be taken into account, as done in fact in Ref.~\cite{Grunhaupt.2018} for resonators. For such devices, one can set $\Gamma_\mathrm{ex} = \Gamma_0 \bar{n}$, where $\bar{n}$ is the average number of photons stored in the resonator and $\Gamma_0$ is in general a material- and geometry-dependent parameter; it can be related to the coupling strength $c^\mathrm{QP}_\mathrm{phot}$~\cite{Fischer.2023} between the photons and the quasiparticles, $\Gamma_0 = c^\mathrm{QP}_\mathrm{phot}\sqrt{2\Delta/\omega_0}$ (here we assume $\omega_0\gg \eta^{2/3}E_g$); 
for aluminum resonators, we estimate $\Gamma_0 \approx 10^{-1}$-$10^{-2}\,$s$^{-1}$. As temperature increases above $\epsilon_T$ the photon absorption rate should be compared to that of thermal phonons,
\begin{equation}\label{eq:Gex}
    \Gamma_\mathrm{ex}^{T} \simeq a_T\, r \left(\frac{T}{\Delta}\right)^{7/2} 
\end{equation}
with $a_T \simeq 0.76$ for $\epsilon_T \ll T \ll \Delta \eta^{2/3}$ and $a_T \simeq 0.66$ for $\Delta \eta^{2/3} \ll T \ll \Delta$ (the latter value is as for BCS superconductors~\cite{Kaplan.1976}).
Note that in writing Eq.~\eqref{eq:gammaloc} we assumed $T \ll \epsilon_T$; in the regime $\epsilon_T \ll T \ll \Delta\eta^{2/3}$, we have $\Gamma_\mathrm{loc}\ll \Gamma^T_\mathrm{ex}$~\cite{supplemental} and hence $\beta\ll\tilde{\beta}$, implying $x_l\ll x_m$, cf. Eq.~\eqref{eq:xmxlbetatilde}.

The model in Eqs.~\eqref{eq: Effective Rate Equation} and \eqref{eq:xm rate eq} makes it also possible to calculate the relaxation rates $\lambda_i$, $i=1,\,2$, of the quasiparticle densities towards the steady state, $x_\alpha(t) \sim \bar{x}_\alpha+\delta x_{\alpha1}(0) e^{-\lambda_1 t}+\delta x_{\alpha2}(0) e^{-\lambda_2 t}$, where $\bar{x}_\alpha$, $\alpha=l,\,m$, are the steady-state densities and $\delta x_{\alpha i}(0)$ account for deviation from the steady state at time $t=0$. Assuming $\kappa>\kappa_c$ so that all the recombination rates take the value $r$, by linearization around the steady state we find $\lambda_1=2 r \bar{x}$ and $\lambda_2 = \Gamma_\mathrm{loc}+\Gamma_\mathrm{ex}+r \bar{x}$~\cite{supplemental}. The rate $\lambda_1$ coincides with that of the original RT model and represents the relaxation of the total density $x$, while $\lambda_2$ is the decay rate of the ``differential'' mode for which $\delta x_l = -\delta x_m$; that is, $\lambda_2$ governs the return to the steady state when quasiparticles are exchanged between localized and mobile states rather than added to the system by a pair-breaking mechanism.

Having obtained estimates for all the relevant rates, we can now addressed the question of which regime is relevant for current experiments in aluminum devices. 
The relevant material parameters are the recombination coefficient $r$ and the localization rate $\Gamma_\mathrm{loc}$, while the generation rate depends on external factors such as temperature, device shielding, and filtering. The recombination coefficient in aluminum is relatively well-known;
for our estimates in thin-film devices we set $r\sim 10^7\,$s$^{-1}$~\cite{Wang.2014}. To evaluate $\Gamma_\mathrm{loc}$, Eq.~\eqref{eq:gammaloc}, we need first to estimate the energy scale $\epsilon_T$; to this end, we note that it must be small compared to the broadening $\Delta - E_g \sim \eta^{2/3}\Delta$ of the peak in the DoS. 
The latter can be extracted from experiments in which the phenomenological Dynes parameter $\gamma_D$ is used in fitting data sensitive to the peak shape using the formula $\rho_D(E) = \mathrm{Re}[(E/\Delta+i\gamma_D)/\sqrt{(E/\Delta+i\gamma_D)^2-1}]$. Although this formula predicts a finite DoS of order $\gamma_D$ also at small energies $E\ll \Delta$, in general the peak broadening and the deep subgap DoS are determined by different parameters; for example, if the broadening is due to the superconductor being in tunnel contact with a normal-metal film, the formula stops being valid below a so-called minigap energy whose value depends on the property of the normal layer~\cite{Hosseinkhani.2018}. For this reason, we do not estimate $\gamma_D$ from measurements sensitive to deep subgap states. Current-voltage measurements in relatively thick (500~nm) Al SINIS structures found $\gamma_D \simeq 3\times10^{-4}$~\cite{ONeil.2010}; experiments on thermal transport~\cite{NNano.2015,NNano.2017} and thermoelectric effect~\cite{NNano.2022} in Josephson junctions with Al films of thicknesses between 14 and 40~nm report $\gamma_D$ between $5\times 10^{-4}$ and $5\times 10^{-3}$. As mentioned above, for the theoretical modeling of the subgap states that we use to be consistent, the condition $\epsilon_T \ll \gamma_D \Delta$ must be satisfied, so we conservatively assume $\epsilon_T/\Delta \sim 5\times10^{-4}$, a value an order of magnitude larger than estimated in Ref.~\cite{Silva.2005}. Then according to Eq.~\eqref{eq:xl_ub} we have $x_l \lesssim \Gamma_\mathrm{loc}/r < 10^{-11}$. Since in Al we have $n_\mathrm{Cp} \simeq 5\times 10^6\,\mu\textrm{m}^3$, this normalized density corresponds to less than one quasiparticle even in large devices such as 3D transmons~\cite{Paik.2011} or coplanar waveguide resonators~\cite{Visser.2014} with volume of the order a few times $10^3\,\mu\textrm{m}^3$. Therefore we must conclude that localization does not contribute to excess quasiparticles in typical aluminum devices. 
Given our estimate above for $\epsilon_T$, we additionally note that in typical experiments at $T \simeq 10\,$mK, we have  $T/\epsilon_T \sim 8$; even though the excitation rate in Eq.~\eqref{eq:Gex} is slow, $\Gamma_\mathrm{ex} \sim 0.03\,$Hz, in rough agreement with the estimate in Ref.~\cite{Silva.2005}, it is nonetheless much faster than both the zero-temperature localization rate from Eq.~\eqref{eq:gammaloc} and its finite temperature counterpart; therefore, we expect that for $T\gtrsim \epsilon_T$ excitation prevails over localization, again pointing to the conclusion that localization is not relevant in determining the quasiparticle density. 
This conclusion also applies to devices made with $\beta$-Ta films, since both $T_c\sim 0.87$~K and $\epsilon_T/\Delta \sim 1.5 \times 10^{-4}$ have values similar to that of Al, according to a recent work~\cite{deRooij.2024}. Interestingly, the density measured there follows the thermal equilibrium expectation, while the decay rate is argued to be the excitation rate $\Gamma_\mathrm{ex}^T$ and therefore compatible with that of the differential mode~\cite{supplemental}.

In the preceding paragraph we have considered typical thin Al films. Even keeping aluminum as the main superconducting material, the value of $\gamma_D$ and hence possibly the ratio $\epsilon_T/\Delta$ can be increased in several ways, for example by Mn doping~\cite{ONeil.2010} and by proximity effect with a normal metal~\cite{Hosseinkhani.2018} such as Cu~\cite{NNano.2022}, methods that however suppress the gap $\Delta$. Alternatively, oxygen doping can increase $\Delta$~\cite{Dynes.1981,Valenti.2019} (up to a point) as well as $\gamma_D$~\cite{Dynes.1984}; in fact, the quasiparticle dynamics in granular aluminum resonators appears to be affected by localization~\cite{Grunhaupt.2018}. Similar behavior has been recently reported in resonators incorporating tungsten silicide~\cite{Larson.2025}. Other materials, such as Nb ($\gamma_D\sim 5\times10^{-2}$~\cite{Nevala.2012}), NbN ($\gamma_D\sim 2.4$-$4\times10^{-2}$~\cite{Chaudhuri.2013}), or TaN ($\gamma_D\sim 7\times10^{-2}$~\cite{Chaudhuri.2014}), also have larger broadening (in units of $\Delta$) than Al. Increasing $\epsilon_T/\Delta$ by an order of magnitude would push up the bound on $x_l$ by more than three orders of magnitude, and also make possible experiments in the regime $T \lesssim \epsilon_T$ in which excitation by thermal phonon is slower than localization. Moreover, materials with higher gap generally have faster recombination (larger $r$) than Al~\cite{Kaplan.1976}, making it easier to reach the regime $\beta > 1$ for a given generation rate $g$, since $\beta \simeq (\epsilon_T/\Delta)^{7/2}\sqrt{r/g_m}$. Therefore, exploring devices fabricated with properly chosen materials could shed further light on the role of subgap states in their performances. On the theoretical side, our considerations could be extended to cleaner materials by building on the approach presented in Ref.~\cite{Graaf.2020}. 
In a complementary direction, one could also consider the case of more strongly coupled magnetic impurities, such that an impurity band of localized states is formed with energy well below the gap~\cite{Fominov.2016}; then localization and mobile-localized quasiparticle recombination could be enhanced~\cite{Kozorezov.2008}, effects that have been used to interpret experiments in superconducting tunnel junction detectors~\cite{Hijmering.2009}.

\paragraph*{Note added:} during the completion of this work, similar conclusions were reached in Ref.~\cite{Ullom} based on current-voltage measurements in Al- and Nb-based junctions.

\paragraph*{Acknowledgments.} We thank G. Marchegani and J. Ullom for interesting discussions. This work was supported in part by the German Federal Ministry of Education and Research (BMBF), funding program ``Quantum Technologies – From Basic Research
to Market,'' project QSolid (Grant No. 13N16149). G.C. acknowledges partial support by the U.S. Government under ARO grant No. W911NF2210257.

\bibliography{sources_sc}

\begin{thebibliography}{54}%
\makeatletter
\providecommand \@ifxundefined [1]{%
 \@ifx{#1\undefined}
}%
\providecommand \@ifnum [1]{%
 \ifnum #1\expandafter \@firstoftwo
 \else \expandafter \@secondoftwo
 \fi
}%
\providecommand \@ifx [1]{%
 \ifx #1\expandafter \@firstoftwo
 \else \expandafter \@secondoftwo
 \fi
}%
\providecommand \natexlab [1]{#1}%
\providecommand \enquote  [1]{``#1''}%
\providecommand \bibnamefont  [1]{#1}%
\providecommand \bibfnamefont [1]{#1}%
\providecommand \citenamefont [1]{#1}%
\providecommand \href@noop [0]{\@secondoftwo}%
\providecommand \href [0]{\begingroup \@sanitize@url \@href}%
\providecommand \@href[1]{\@@startlink{#1}\@@href}%
\providecommand \@@href[1]{\endgroup#1\@@endlink}%
\providecommand \@sanitize@url [0]{\catcode `\\12\catcode `\$12\catcode
  `\&12\catcode `\#12\catcode `\^12\catcode `\_12\catcode `\%12\relax}%
\providecommand \@@startlink[1]{}%
\providecommand \@@endlink[0]{}%
\providecommand \url  [0]{\begingroup\@sanitize@url \@url }%
\providecommand \@url [1]{\endgroup\@href {#1}{\urlprefix }}%
\providecommand \urlprefix  [0]{URL }%
\providecommand \Eprint [0]{\href }%
\providecommand \doibase [0]{https://doi.org/}%
\providecommand \selectlanguage [0]{\@gobble}%
\providecommand \bibinfo  [0]{\@secondoftwo}%
\providecommand \bibfield  [0]{\@secondoftwo}%
\providecommand \translation [1]{[#1]}%
\providecommand \BibitemOpen [0]{}%
\providecommand \bibitemStop [0]{}%
\providecommand \bibitemNoStop [0]{.\EOS\space}%
\providecommand \EOS [0]{\spacefactor3000\relax}%
\providecommand \BibitemShut  [1]{\csname bibitem#1\endcsname}%
\let\auto@bib@innerbib\@empty
\bibitem [{\citenamefont {Veps\"al\"ainen}\ \emph {et~al.}(2020)\citenamefont
  {Veps\"al\"ainen}, \citenamefont {Karamlou}, \citenamefont {Orrell},
  \citenamefont {Dogra}, \citenamefont {Loer}, \citenamefont {Vasconcelos},
  \citenamefont {Kim}, \citenamefont {Melville}, \citenamefont {Niedzielski},
  \citenamefont {Yoder}, \citenamefont {Gustavsson}, \citenamefont {Formaggio},
  \citenamefont {VanDevender},\ and\ \citenamefont
  {Oliver}}]{Vepsalainen.2020}%
  \BibitemOpen
  \bibfield  {author} {\bibinfo {author} {\bibfnamefont {A.~P.}\ \bibnamefont
  {Veps\"al\"ainen}}, \bibinfo {author} {\bibfnamefont {A.~H.}\ \bibnamefont
  {Karamlou}}, \bibinfo {author} {\bibfnamefont {J.~L.}\ \bibnamefont
  {Orrell}}, \bibinfo {author} {\bibfnamefont {A.~S.}\ \bibnamefont {Dogra}},
  \bibinfo {author} {\bibfnamefont {B.}~\bibnamefont {Loer}}, \bibinfo {author}
  {\bibfnamefont {F.}~\bibnamefont {Vasconcelos}}, \bibinfo {author}
  {\bibfnamefont {D.~K.}\ \bibnamefont {Kim}}, \bibinfo {author} {\bibfnamefont
  {A.~J.}\ \bibnamefont {Melville}}, \bibinfo {author} {\bibfnamefont {B.~M.}\
  \bibnamefont {Niedzielski}}, \bibinfo {author} {\bibfnamefont {J.~L.}\
  \bibnamefont {Yoder}}, \bibinfo {author} {\bibfnamefont {S.}~\bibnamefont
  {Gustavsson}}, \bibinfo {author} {\bibfnamefont {J.~A.}\ \bibnamefont
  {Formaggio}}, \bibinfo {author} {\bibfnamefont {B.~A.}\ \bibnamefont
  {VanDevender}},\ and\ \bibinfo {author} {\bibfnamefont {W.~D.}\ \bibnamefont
  {Oliver}},\ }\bibfield  {title} {\bibinfo {title} {Impact of ionizing
  radiation on superconducting qubit coherence},\ }\href
  {https://doi.org/10.1038/s41586-020-2619-8} {\bibfield  {journal} {\bibinfo
  {journal} {Nature}\ }\textbf {\bibinfo {volume} {584}},\ \bibinfo {pages}
  {551} (\bibinfo {year} {2020})}\BibitemShut {NoStop}%
\bibitem [{\citenamefont {Cardani}\ \emph {et~al.}(2021)\citenamefont
  {Cardani}, \citenamefont {Valenti}, \citenamefont {Casali}, \citenamefont
  {Catelani}, \citenamefont {Charpentier}, \citenamefont {Clemenza},
  \citenamefont {Colantoni}, \citenamefont {Cruciani}, \citenamefont
  {D'Imperio}, \citenamefont {Gironi}, \citenamefont {Gr{\"u}nhaupt},
  \citenamefont {Gusenkova}, \citenamefont {Henriques}, \citenamefont {Lagoin},
  \citenamefont {Martinez}, \citenamefont {Pettinari}, \citenamefont {Rusconi},
  \citenamefont {Sander}, \citenamefont {Tomei}, \citenamefont {Ustinov},
  \citenamefont {Weber}, \citenamefont {Wernsdorfer}, \citenamefont {Vignati},
  \citenamefont {Pirro},\ and\ \citenamefont {Pop}}]{Cardani.2021}%
  \BibitemOpen
  \bibfield  {author} {\bibinfo {author} {\bibfnamefont {L.}~\bibnamefont
  {Cardani}}, \bibinfo {author} {\bibfnamefont {F.}~\bibnamefont {Valenti}},
  \bibinfo {author} {\bibfnamefont {N.}~\bibnamefont {Casali}}, \bibinfo
  {author} {\bibfnamefont {G.}~\bibnamefont {Catelani}}, \bibinfo {author}
  {\bibfnamefont {T.}~\bibnamefont {Charpentier}}, \bibinfo {author}
  {\bibfnamefont {M.}~\bibnamefont {Clemenza}}, \bibinfo {author}
  {\bibfnamefont {I.}~\bibnamefont {Colantoni}}, \bibinfo {author}
  {\bibfnamefont {A.}~\bibnamefont {Cruciani}}, \bibinfo {author}
  {\bibfnamefont {G.}~\bibnamefont {D'Imperio}}, \bibinfo {author}
  {\bibfnamefont {L.}~\bibnamefont {Gironi}}, \bibinfo {author} {\bibfnamefont
  {L.}~\bibnamefont {Gr{\"u}nhaupt}}, \bibinfo {author} {\bibfnamefont
  {D.}~\bibnamefont {Gusenkova}}, \bibinfo {author} {\bibfnamefont
  {F.}~\bibnamefont {Henriques}}, \bibinfo {author} {\bibfnamefont
  {M.}~\bibnamefont {Lagoin}}, \bibinfo {author} {\bibfnamefont
  {M.}~\bibnamefont {Martinez}}, \bibinfo {author} {\bibfnamefont
  {G.}~\bibnamefont {Pettinari}}, \bibinfo {author} {\bibfnamefont
  {C.}~\bibnamefont {Rusconi}}, \bibinfo {author} {\bibfnamefont
  {O.}~\bibnamefont {Sander}}, \bibinfo {author} {\bibfnamefont
  {C.}~\bibnamefont {Tomei}}, \bibinfo {author} {\bibfnamefont {A.~V.}\
  \bibnamefont {Ustinov}}, \bibinfo {author} {\bibfnamefont {M.}~\bibnamefont
  {Weber}}, \bibinfo {author} {\bibfnamefont {W.}~\bibnamefont {Wernsdorfer}},
  \bibinfo {author} {\bibfnamefont {M.}~\bibnamefont {Vignati}}, \bibinfo
  {author} {\bibfnamefont {S.}~\bibnamefont {Pirro}},\ and\ \bibinfo {author}
  {\bibfnamefont {I.~M.}\ \bibnamefont {Pop}},\ }\bibfield  {title} {\bibinfo
  {title} {Reducing the impact of radioactivity on quantum circuits in a
  deep-underground facility},\ }\href
  {https://doi.org/10.1038/s41467-021-23032-z} {\bibfield  {journal} {\bibinfo
  {journal} {Nat. Commun.}\ }\textbf {\bibinfo {volume} {12}},\ \bibinfo
  {pages} {2733} (\bibinfo {year} {2021})}\BibitemShut {NoStop}%
\bibitem [{\citenamefont {Wilen}\ \emph {et~al.}(2021)\citenamefont {Wilen},
  \citenamefont {Abdullah}, \citenamefont {Kurinsky}, \citenamefont {Stanford},
  \citenamefont {Cardani}, \citenamefont {D’Imperio}, \citenamefont {Tomei},
  \citenamefont {Faoro}, \citenamefont {Ioffe}, \citenamefont {Liu},
  \citenamefont {Opremcak}, \citenamefont {Christensen}, \citenamefont
  {DuBois},\ and\ \citenamefont {McDermott}}]{Wilen.2021}%
  \BibitemOpen
  \bibfield  {author} {\bibinfo {author} {\bibfnamefont {C.~D.}\ \bibnamefont
  {Wilen}}, \bibinfo {author} {\bibfnamefont {S.}~\bibnamefont {Abdullah}},
  \bibinfo {author} {\bibfnamefont {N.~A.}\ \bibnamefont {Kurinsky}}, \bibinfo
  {author} {\bibfnamefont {C.}~\bibnamefont {Stanford}}, \bibinfo {author}
  {\bibfnamefont {L.}~\bibnamefont {Cardani}}, \bibinfo {author} {\bibfnamefont
  {G.}~\bibnamefont {D’Imperio}}, \bibinfo {author} {\bibfnamefont
  {C.}~\bibnamefont {Tomei}}, \bibinfo {author} {\bibfnamefont
  {L.}~\bibnamefont {Faoro}}, \bibinfo {author} {\bibfnamefont {L.~B.}\
  \bibnamefont {Ioffe}}, \bibinfo {author} {\bibfnamefont {C.~H.}\ \bibnamefont
  {Liu}}, \bibinfo {author} {\bibfnamefont {A.}~\bibnamefont {Opremcak}},
  \bibinfo {author} {\bibfnamefont {B.~G.}\ \bibnamefont {Christensen}},
  \bibinfo {author} {\bibfnamefont {J.~L.}\ \bibnamefont {DuBois}},\ and\
  \bibinfo {author} {\bibfnamefont {R.}~\bibnamefont {McDermott}},\ }\bibfield
  {title} {\bibinfo {title} {Correlated charge noise and relaxation errors in
  superconducting qubits},\ }\href {https://doi.org/10.1038/s41586-021-03557-5}
  {\bibfield  {journal} {\bibinfo  {journal} {Nature}\ }\textbf {\bibinfo
  {volume} {594}},\ \bibinfo {pages} {369} (\bibinfo {year}
  {2021})}\BibitemShut {NoStop}%
\bibitem [{\citenamefont {Houzet}\ \emph {et~al.}(2019)\citenamefont {Houzet},
  \citenamefont {Serniak}, \citenamefont {Catelani}, \citenamefont {Devoret},\
  and\ \citenamefont {Glazman}}]{Houzet.2019}%
  \BibitemOpen
  \bibfield  {author} {\bibinfo {author} {\bibfnamefont {M.}~\bibnamefont
  {Houzet}}, \bibinfo {author} {\bibfnamefont {K.}~\bibnamefont {Serniak}},
  \bibinfo {author} {\bibfnamefont {G.}~\bibnamefont {Catelani}}, \bibinfo
  {author} {\bibfnamefont {M.~H.}\ \bibnamefont {Devoret}},\ and\ \bibinfo
  {author} {\bibfnamefont {L.~I.}\ \bibnamefont {Glazman}},\ }\bibfield
  {title} {\bibinfo {title} {Photon-assisted charge-parity jumps in a
  superconducting qubit},\ }\href
  {https://doi.org/10.1103/PhysRevLett.123.107704} {\bibfield  {journal}
  {\bibinfo  {journal} {Phys. Rev. Lett.}\ }\textbf {\bibinfo {volume} {123}},\
  \bibinfo {pages} {107704} (\bibinfo {year} {2019})}\BibitemShut {NoStop}%
\bibitem [{\citenamefont {Diamond}\ \emph {et~al.}(2022)\citenamefont
  {Diamond}, \citenamefont {Fatemi}, \citenamefont {Hays}, \citenamefont {Nho},
  \citenamefont {Kurilovich}, \citenamefont {Connolly}, \citenamefont {Joshi},
  \citenamefont {Serniak}, \citenamefont {Frunzio}, \citenamefont {Glazman},\
  and\ \citenamefont {Devoret}}]{Diamond.2022}%
  \BibitemOpen
  \bibfield  {author} {\bibinfo {author} {\bibfnamefont {S.}~\bibnamefont
  {Diamond}}, \bibinfo {author} {\bibfnamefont {V.}~\bibnamefont {Fatemi}},
  \bibinfo {author} {\bibfnamefont {M.}~\bibnamefont {Hays}}, \bibinfo {author}
  {\bibfnamefont {H.}~\bibnamefont {Nho}}, \bibinfo {author} {\bibfnamefont
  {P.~D.}\ \bibnamefont {Kurilovich}}, \bibinfo {author} {\bibfnamefont
  {T.}~\bibnamefont {Connolly}}, \bibinfo {author} {\bibfnamefont {V.~R.}\
  \bibnamefont {Joshi}}, \bibinfo {author} {\bibfnamefont {K.}~\bibnamefont
  {Serniak}}, \bibinfo {author} {\bibfnamefont {L.}~\bibnamefont {Frunzio}},
  \bibinfo {author} {\bibfnamefont {L.~I.}\ \bibnamefont {Glazman}},\ and\
  \bibinfo {author} {\bibfnamefont {M.~H.}\ \bibnamefont {Devoret}},\
  }\bibfield  {title} {\bibinfo {title} {Distinguishing parity-switching
  mechanisms in a superconducting qubit},\ }\href
  {https://doi.org/10.1103/PRXQuantum.3.040304} {\bibfield  {journal} {\bibinfo
   {journal} {PRX Quantum}\ }\textbf {\bibinfo {volume} {3}},\ \bibinfo {pages}
  {040304} (\bibinfo {year} {2022})}\BibitemShut {NoStop}%
\bibitem [{\citenamefont {Pan}\ \emph {et~al.}(2022)\citenamefont {Pan},
  \citenamefont {Zhou}, \citenamefont {Yuan}, \citenamefont {Nie},
  \citenamefont {Wei}, \citenamefont {Zhang}, \citenamefont {Li}, \citenamefont
  {Liu}, \citenamefont {Jiang}, \citenamefont {Catelani}, \citenamefont {Hu},
  \citenamefont {Yan},\ and\ \citenamefont {Yu}}]{Pan.2022}%
  \BibitemOpen
  \bibfield  {author} {\bibinfo {author} {\bibfnamefont {X.}~\bibnamefont
  {Pan}}, \bibinfo {author} {\bibfnamefont {Y.}~\bibnamefont {Zhou}}, \bibinfo
  {author} {\bibfnamefont {H.}~\bibnamefont {Yuan}}, \bibinfo {author}
  {\bibfnamefont {L.}~\bibnamefont {Nie}}, \bibinfo {author} {\bibfnamefont
  {W.}~\bibnamefont {Wei}}, \bibinfo {author} {\bibfnamefont {L.}~\bibnamefont
  {Zhang}}, \bibinfo {author} {\bibfnamefont {J.}~\bibnamefont {Li}}, \bibinfo
  {author} {\bibfnamefont {S.}~\bibnamefont {Liu}}, \bibinfo {author}
  {\bibfnamefont {Z.~H.}\ \bibnamefont {Jiang}}, \bibinfo {author}
  {\bibfnamefont {G.}~\bibnamefont {Catelani}}, \bibinfo {author}
  {\bibfnamefont {L.}~\bibnamefont {Hu}}, \bibinfo {author} {\bibfnamefont
  {F.}~\bibnamefont {Yan}},\ and\ \bibinfo {author} {\bibfnamefont
  {D.}~\bibnamefont {Yu}},\ }\bibfield  {title} {\bibinfo {title} {Engineering
  superconducting qubits to reduce quasiparticles and charge noise},\ }\href
  {https://doi.org/10.1038/s41467-022-34727-2} {\bibfield  {journal} {\bibinfo
  {journal} {Nat. Commun.}\ }\textbf {\bibinfo {volume} {13}},\ \bibinfo
  {pages} {7196} (\bibinfo {year} {2022})}\BibitemShut {NoStop}%
\bibitem [{\citenamefont {Liu}\ \emph {et~al.}(2024)\citenamefont {Liu},
  \citenamefont {Harrison}, \citenamefont {Patel}, \citenamefont {Wilen},
  \citenamefont {Rafferty}, \citenamefont {Shearrow}, \citenamefont {Ballard},
  \citenamefont {Iaia}, \citenamefont {Ku}, \citenamefont {Plourde},\ and\
  \citenamefont {McDermott}}]{Liu.2024}%
  \BibitemOpen
  \bibfield  {author} {\bibinfo {author} {\bibfnamefont {C.~H.}\ \bibnamefont
  {Liu}}, \bibinfo {author} {\bibfnamefont {D.~C.}\ \bibnamefont {Harrison}},
  \bibinfo {author} {\bibfnamefont {S.}~\bibnamefont {Patel}}, \bibinfo
  {author} {\bibfnamefont {C.~D.}\ \bibnamefont {Wilen}}, \bibinfo {author}
  {\bibfnamefont {O.}~\bibnamefont {Rafferty}}, \bibinfo {author}
  {\bibfnamefont {A.}~\bibnamefont {Shearrow}}, \bibinfo {author}
  {\bibfnamefont {A.}~\bibnamefont {Ballard}}, \bibinfo {author} {\bibfnamefont
  {V.}~\bibnamefont {Iaia}}, \bibinfo {author} {\bibfnamefont {J.}~\bibnamefont
  {Ku}}, \bibinfo {author} {\bibfnamefont {B.~L.~T.}\ \bibnamefont {Plourde}},\
  and\ \bibinfo {author} {\bibfnamefont {R.}~\bibnamefont {McDermott}},\
  }\bibfield  {title} {\bibinfo {title} {Quasiparticle poisoning of
  superconducting qubits from resonant absorption of pair-breaking photons},\
  }\href {https://doi.org/10.1103/PhysRevLett.132.017001} {\bibfield  {journal}
  {\bibinfo  {journal} {Phys. Rev. Lett.}\ }\textbf {\bibinfo {volume} {132}},\
  \bibinfo {pages} {017001} (\bibinfo {year} {2024})}\BibitemShut {NoStop}%
\bibitem [{\citenamefont {Anthony-Petersen}\ \emph {et~al.}(2024)\citenamefont
  {Anthony-Petersen}, \citenamefont {Biekert}, \citenamefont {Bunker},
  \citenamefont {Chang}, \citenamefont {Chang}, \citenamefont {Chaplinsky},
  \citenamefont {Fascione}, \citenamefont {Fink}, \citenamefont
  {Garcia-Sciveres}, \citenamefont {Germond}, \citenamefont {Guo},
  \citenamefont {Hertel}, \citenamefont {Hong}, \citenamefont {Kurinsky},
  \citenamefont {Li}, \citenamefont {Lin}, \citenamefont {Lisovenko},
  \citenamefont {Mahapatra}, \citenamefont {Mayer}, \citenamefont {McKinsey},
  \citenamefont {Mehrotra}, \citenamefont {Mirabolfathi}, \citenamefont
  {Neblosky}, \citenamefont {Page}, \citenamefont {Patel}, \citenamefont
  {Penning}, \citenamefont {Pinckney}, \citenamefont {Platt}, \citenamefont
  {Pyle}, \citenamefont {Reed}, \citenamefont {Romani}, \citenamefont
  {Queiroz}, \citenamefont {Sadoulet}, \citenamefont {Serfass}, \citenamefont
  {Smith}, \citenamefont {Sorensen}, \citenamefont {Suerfu}, \citenamefont
  {Suzuki}, \citenamefont {Underwood}, \citenamefont {Velan}, \citenamefont
  {Wang}, \citenamefont {Wang}, \citenamefont {Watkins}, \citenamefont
  {Williams}, \citenamefont {Yefremenko},\ and\ \citenamefont
  {Zhang}}]{RAP.2024}%
  \BibitemOpen
  \bibfield  {author} {\bibinfo {author} {\bibfnamefont {R.}~\bibnamefont
  {Anthony-Petersen}}, \bibinfo {author} {\bibfnamefont {A.}~\bibnamefont
  {Biekert}}, \bibinfo {author} {\bibfnamefont {R.}~\bibnamefont {Bunker}},
  \bibinfo {author} {\bibfnamefont {C.~L.}\ \bibnamefont {Chang}}, \bibinfo
  {author} {\bibfnamefont {Y.-Y.}\ \bibnamefont {Chang}}, \bibinfo {author}
  {\bibfnamefont {L.}~\bibnamefont {Chaplinsky}}, \bibinfo {author}
  {\bibfnamefont {E.}~\bibnamefont {Fascione}}, \bibinfo {author}
  {\bibfnamefont {C.~W.}\ \bibnamefont {Fink}}, \bibinfo {author}
  {\bibfnamefont {M.}~\bibnamefont {Garcia-Sciveres}}, \bibinfo {author}
  {\bibfnamefont {R.}~\bibnamefont {Germond}}, \bibinfo {author} {\bibfnamefont
  {W.}~\bibnamefont {Guo}}, \bibinfo {author} {\bibfnamefont {S.~A.}\
  \bibnamefont {Hertel}}, \bibinfo {author} {\bibfnamefont {Z.}~\bibnamefont
  {Hong}}, \bibinfo {author} {\bibfnamefont {N.}~\bibnamefont {Kurinsky}},
  \bibinfo {author} {\bibfnamefont {X.}~\bibnamefont {Li}}, \bibinfo {author}
  {\bibfnamefont {J.}~\bibnamefont {Lin}}, \bibinfo {author} {\bibfnamefont
  {M.}~\bibnamefont {Lisovenko}}, \bibinfo {author} {\bibfnamefont
  {R.}~\bibnamefont {Mahapatra}}, \bibinfo {author} {\bibfnamefont
  {A.}~\bibnamefont {Mayer}}, \bibinfo {author} {\bibfnamefont {D.~N.}\
  \bibnamefont {McKinsey}}, \bibinfo {author} {\bibfnamefont {S.}~\bibnamefont
  {Mehrotra}}, \bibinfo {author} {\bibfnamefont {N.}~\bibnamefont
  {Mirabolfathi}}, \bibinfo {author} {\bibfnamefont {B.}~\bibnamefont
  {Neblosky}}, \bibinfo {author} {\bibfnamefont {W.~A.}\ \bibnamefont {Page}},
  \bibinfo {author} {\bibfnamefont {P.~K.}\ \bibnamefont {Patel}}, \bibinfo
  {author} {\bibfnamefont {B.}~\bibnamefont {Penning}}, \bibinfo {author}
  {\bibfnamefont {H.~D.}\ \bibnamefont {Pinckney}}, \bibinfo {author}
  {\bibfnamefont {M.}~\bibnamefont {Platt}}, \bibinfo {author} {\bibfnamefont
  {M.}~\bibnamefont {Pyle}}, \bibinfo {author} {\bibfnamefont {M.}~\bibnamefont
  {Reed}}, \bibinfo {author} {\bibfnamefont {R.~K.}\ \bibnamefont {Romani}},
  \bibinfo {author} {\bibfnamefont {H.~S.}\ \bibnamefont {Queiroz}}, \bibinfo
  {author} {\bibfnamefont {B.}~\bibnamefont {Sadoulet}}, \bibinfo {author}
  {\bibfnamefont {B.}~\bibnamefont {Serfass}}, \bibinfo {author} {\bibfnamefont
  {R.}~\bibnamefont {Smith}}, \bibinfo {author} {\bibfnamefont
  {P.}~\bibnamefont {Sorensen}}, \bibinfo {author} {\bibfnamefont
  {B.}~\bibnamefont {Suerfu}}, \bibinfo {author} {\bibfnamefont
  {A.}~\bibnamefont {Suzuki}}, \bibinfo {author} {\bibfnamefont
  {R.}~\bibnamefont {Underwood}}, \bibinfo {author} {\bibfnamefont
  {V.}~\bibnamefont {Velan}}, \bibinfo {author} {\bibfnamefont
  {G.}~\bibnamefont {Wang}}, \bibinfo {author} {\bibfnamefont {Y.}~\bibnamefont
  {Wang}}, \bibinfo {author} {\bibfnamefont {S.~L.}\ \bibnamefont {Watkins}},
  \bibinfo {author} {\bibfnamefont {M.~R.}\ \bibnamefont {Williams}}, \bibinfo
  {author} {\bibfnamefont {V.}~\bibnamefont {Yefremenko}},\ and\ \bibinfo
  {author} {\bibfnamefont {J.}~\bibnamefont {Zhang}},\ }\bibfield  {title}
  {\bibinfo {title} {A stress-induced source of phonon bursts and quasiparticle
  poisoning},\ }\href {https://doi.org/10.1038/s41467-024-50173-8} {\bibfield
  {journal} {\bibinfo  {journal} {Nat. Commun.}\ }\textbf {\bibinfo {volume}
  {15}},\ \bibinfo {pages} {6444} (\bibinfo {year} {2024})}\BibitemShut
  {NoStop}%
\bibitem [{\citenamefont {Yelton}\ \emph {et~al.}(2025)\citenamefont {Yelton},
  \citenamefont {Larson}, \citenamefont {Dodge}, \citenamefont {Okubo},\ and\
  \citenamefont {Plourde}}]{Yelton.2025}%
  \BibitemOpen
  \bibfield  {author} {\bibinfo {author} {\bibfnamefont {E.}~\bibnamefont
  {Yelton}}, \bibinfo {author} {\bibfnamefont {C.~P.}\ \bibnamefont {Larson}},
  \bibinfo {author} {\bibfnamefont {K.}~\bibnamefont {Dodge}}, \bibinfo
  {author} {\bibfnamefont {K.}~\bibnamefont {Okubo}},\ and\ \bibinfo {author}
  {\bibfnamefont {B.~L.~T.}\ \bibnamefont {Plourde}},\ }\bibfield  {title}
  {\bibinfo {title} {Correlated quasiparticle poisoning from phonon-only events
  in superconducting qubits},\ }\href {https://arxiv.org/abs/2503.09554}
  {\bibfield  {journal} {\bibinfo  {journal} {arXiv:2503.09554}\ } (\bibinfo
  {year} {2025})}\BibitemShut {NoStop}%
\bibitem [{\citenamefont {Rothwarf}\ and\ \citenamefont
  {Taylor}(1967)}]{Rothwarf.1967}%
  \BibitemOpen
  \bibfield  {author} {\bibinfo {author} {\bibfnamefont {A.}~\bibnamefont
  {Rothwarf}}\ and\ \bibinfo {author} {\bibfnamefont {B.~N.}\ \bibnamefont
  {Taylor}},\ }\bibfield  {title} {\bibinfo {title} {Measurement of
  recombination lifetimes in superconductors},\ }\href
  {https://doi.org/10.1103/PhysRevLett.19.27} {\bibfield  {journal} {\bibinfo
  {journal} {Phys. Rev. Lett.}\ }\textbf {\bibinfo {volume} {19}},\ \bibinfo
  {pages} {27} (\bibinfo {year} {1967})}\BibitemShut {NoStop}%
\bibitem [{\citenamefont {Bespalov}\ \emph {et~al.}(2016)\citenamefont
  {Bespalov}, \citenamefont {Houzet}, \citenamefont {Meyer},\ and\
  \citenamefont {Nazarov}}]{Bespalov.2016}%
  \BibitemOpen
  \bibfield  {author} {\bibinfo {author} {\bibfnamefont {A.}~\bibnamefont
  {Bespalov}}, \bibinfo {author} {\bibfnamefont {M.}~\bibnamefont {Houzet}},
  \bibinfo {author} {\bibfnamefont {J.~S.}\ \bibnamefont {Meyer}},\ and\
  \bibinfo {author} {\bibfnamefont {Y.~V.}\ \bibnamefont {Nazarov}},\
  }\bibfield  {title} {\bibinfo {title} {Theoretical model to explain excess of
  quasiparticles in superconductors},\ }\href
  {https://doi.org/10.1103/PhysRevLett.117.117002} {\bibfield  {journal}
  {\bibinfo  {journal} {Phys. Rev. Lett.}\ }\textbf {\bibinfo {volume} {117}},\
  \bibinfo {pages} {117002} (\bibinfo {year} {2016})}\BibitemShut {NoStop}%
\bibitem [{\citenamefont {Gr{\"u}nhaupt}\ \emph {et~al.}(2018)\citenamefont
  {Gr{\"u}nhaupt}, \citenamefont {Maleeva}, \citenamefont {Skacel},
  \citenamefont {Calvo}, \citenamefont {Levy-Bertrand}, \citenamefont
  {Ustinov}, \citenamefont {Rotzinger}, \citenamefont {Monfardini},
  \citenamefont {Catelani},\ and\ \citenamefont {Pop}}]{Grunhaupt.2018}%
  \BibitemOpen
  \bibfield  {author} {\bibinfo {author} {\bibfnamefont {L.}~\bibnamefont
  {Gr{\"u}nhaupt}}, \bibinfo {author} {\bibfnamefont {N.}~\bibnamefont
  {Maleeva}}, \bibinfo {author} {\bibfnamefont {S.~T.}\ \bibnamefont {Skacel}},
  \bibinfo {author} {\bibfnamefont {M.}~\bibnamefont {Calvo}}, \bibinfo
  {author} {\bibfnamefont {F.}~\bibnamefont {Levy-Bertrand}}, \bibinfo {author}
  {\bibfnamefont {A.~V.}\ \bibnamefont {Ustinov}}, \bibinfo {author}
  {\bibfnamefont {H.}~\bibnamefont {Rotzinger}}, \bibinfo {author}
  {\bibfnamefont {A.}~\bibnamefont {Monfardini}}, \bibinfo {author}
  {\bibfnamefont {G.}~\bibnamefont {Catelani}},\ and\ \bibinfo {author}
  {\bibfnamefont {I.~M.}\ \bibnamefont {Pop}},\ }\bibfield  {title} {\bibinfo
  {title} {Loss mechanisms and quasiparticle dynamics in superconducting
  microwave resonators made of thin-film granular aluminum},\ }\href
  {https://doi.org/10.1103/PhysRevLett.121.117001} {\bibfield  {journal}
  {\bibinfo  {journal} {Phys. Rev. Lett.}\ }\textbf {\bibinfo {volume} {121}},\
  \bibinfo {pages} {117001} (\bibinfo {year} {2018})}\BibitemShut {NoStop}%
\bibitem [{\citenamefont {Sac\'ep\'e}\ \emph {et~al.}(2008)\citenamefont
  {Sac\'ep\'e}, \citenamefont {Chapelier}, \citenamefont {Baturina},
  \citenamefont {Vinokur}, \citenamefont {Baklanov},\ and\ \citenamefont
  {Sanquer}}]{Sacepe.2008}%
  \BibitemOpen
  \bibfield  {author} {\bibinfo {author} {\bibfnamefont {B.}~\bibnamefont
  {Sac\'ep\'e}}, \bibinfo {author} {\bibfnamefont {C.}~\bibnamefont
  {Chapelier}}, \bibinfo {author} {\bibfnamefont {T.~I.}\ \bibnamefont
  {Baturina}}, \bibinfo {author} {\bibfnamefont {V.~M.}\ \bibnamefont
  {Vinokur}}, \bibinfo {author} {\bibfnamefont {M.~R.}\ \bibnamefont
  {Baklanov}},\ and\ \bibinfo {author} {\bibfnamefont {M.}~\bibnamefont
  {Sanquer}},\ }\bibfield  {title} {\bibinfo {title} {Disorder-induced
  inhomogeneities of the superconducting state close to the
  superconductor-insulator transition},\ }\href
  {https://doi.org/10.1103/PhysRevLett.101.157006} {\bibfield  {journal}
  {\bibinfo  {journal} {Phys. Rev. Lett.}\ }\textbf {\bibinfo {volume} {101}},\
  \bibinfo {pages} {157006} (\bibinfo {year} {2008})}\BibitemShut {NoStop}%
\bibitem [{\citenamefont {Kulik}\ and\ \citenamefont
  {Itskovich}(1968)}]{Kulik.1968}%
  \BibitemOpen
  \bibfield  {author} {\bibinfo {author} {\bibfnamefont {I.~O.}\ \bibnamefont
  {Kulik}}\ and\ \bibinfo {author} {\bibfnamefont {O.~Y.}\ \bibnamefont
  {Itskovich}},\ }\bibfield  {title} {\bibinfo {title} {Effect of concentration
  non-uniformity on the properties of superconductors with paramagnetic
  impurities},\ }\href {http://jetp.ras.ru/cgi-bin/e/index/e/28/1/p102?a=list}
  {\bibfield  {journal} {\bibinfo  {journal} {J. Exp. Theor. Phys.}\ }\textbf
  {\bibinfo {volume} {28}},\ \bibinfo {pages} {102} (\bibinfo {year}
  {1968})}\BibitemShut {NoStop}%
\bibitem [{\citenamefont {Larkin}\ and\ \citenamefont
  {Ovchinnikov}(1972)}]{Larkin.1972}%
  \BibitemOpen
  \bibfield  {author} {\bibinfo {author} {\bibfnamefont {A.~I.}\ \bibnamefont
  {Larkin}}\ and\ \bibinfo {author} {\bibfnamefont {Y.~N.}\ \bibnamefont
  {Ovchinnikov}},\ }\bibfield  {title} {\bibinfo {title} {Density of states in
  inhomogeneous superconductors},\ }\href
  {http://jetp.ras.ru/cgi-bin/e/index/e/34/5/p1144?a=list} {\bibfield
  {journal} {\bibinfo  {journal} {J. Exp. Theor. Phys.}\ }\textbf {\bibinfo
  {volume} {34}},\ \bibinfo {pages} {1144} (\bibinfo {year}
  {1972})}\BibitemShut {NoStop}%
\bibitem [{\citenamefont {Matthias}\ \emph {et~al.}(1958)\citenamefont
  {Matthias}, \citenamefont {Suhl},\ and\ \citenamefont
  {Corenzwit}}]{Matthias.1958}%
  \BibitemOpen
  \bibfield  {author} {\bibinfo {author} {\bibfnamefont {B.~T.}\ \bibnamefont
  {Matthias}}, \bibinfo {author} {\bibfnamefont {H.}~\bibnamefont {Suhl}},\
  and\ \bibinfo {author} {\bibfnamefont {E.}~\bibnamefont {Corenzwit}},\
  }\bibfield  {title} {\bibinfo {title} {Spin exchange in superconductors},\
  }\href {https://doi.org/10.1103/PhysRevLett.1.92} {\bibfield  {journal}
  {\bibinfo  {journal} {Phys. Rev. Lett.}\ }\textbf {\bibinfo {volume} {1}},\
  \bibinfo {pages} {92} (\bibinfo {year} {1958})}\BibitemShut {NoStop}%
\bibitem [{\citenamefont {Abrikosov}\ and\ \citenamefont
  {Gorkov}(1961)}]{AG.1961}%
  \BibitemOpen
  \bibfield  {author} {\bibinfo {author} {\bibfnamefont {A.}~\bibnamefont
  {Abrikosov}}\ and\ \bibinfo {author} {\bibfnamefont {L.}~\bibnamefont
  {Gorkov}},\ }\bibfield  {title} {\bibinfo {title} {Contribution to the theory
  of superconducting alloys with paramagnetic impurities},\ }\href@noop {}
  {\bibfield  {journal} {\bibinfo  {journal} {Soviet Physics JETP}\ }\textbf
  {\bibinfo {volume} {12}},\ \bibinfo {pages} {1243} (\bibinfo {year}
  {1961})}\BibitemShut {NoStop}%
\bibitem [{\citenamefont {Skalski}\ \emph {et~al.}(1964)\citenamefont
  {Skalski}, \citenamefont {Betbeder-Matibet},\ and\ \citenamefont
  {Weiss}}]{Skalski.1964}%
  \BibitemOpen
  \bibfield  {author} {\bibinfo {author} {\bibfnamefont {S.}~\bibnamefont
  {Skalski}}, \bibinfo {author} {\bibfnamefont {O.}~\bibnamefont
  {Betbeder-Matibet}},\ and\ \bibinfo {author} {\bibfnamefont {P.~R.}\
  \bibnamefont {Weiss}},\ }\bibfield  {title} {\bibinfo {title} {Properties of
  superconducting alloys containing paramagnetic impurities},\ }\href
  {https://doi.org/10.1103/PhysRev.136.A1500} {\bibfield  {journal} {\bibinfo
  {journal} {Phys. Rev.}\ }\textbf {\bibinfo {volume} {136}},\ \bibinfo {pages}
  {A1500} (\bibinfo {year} {1964})}\BibitemShut {NoStop}%
\bibitem [{\citenamefont {Silva}\ and\ \citenamefont
  {Ioffe}(2005)}]{Silva.2005}%
  \BibitemOpen
  \bibfield  {author} {\bibinfo {author} {\bibfnamefont {A.}~\bibnamefont
  {Silva}}\ and\ \bibinfo {author} {\bibfnamefont {L.~B.}\ \bibnamefont
  {Ioffe}},\ }\bibfield  {title} {\bibinfo {title} {Subgap states in dirty
  superconductors and their effect on dephasing in {Josephson} qubits},\ }\href
  {https://doi.org/10.1103/PhysRevB.71.104502} {\bibfield  {journal} {\bibinfo
  {journal} {Phys. Rev. B}\ }\textbf {\bibinfo {volume} {71}},\ \bibinfo
  {pages} {104502} (\bibinfo {year} {2005})}\BibitemShut {NoStop}%
\bibitem [{\citenamefont {Skvortsov}\ and\ \citenamefont
  {Feigel'man}(2013)}]{Skvortsov.2013}%
  \BibitemOpen
  \bibfield  {author} {\bibinfo {author} {\bibfnamefont {M.~A.}\ \bibnamefont
  {Skvortsov}}\ and\ \bibinfo {author} {\bibfnamefont {M.~V.}\ \bibnamefont
  {Feigel'man}},\ }\bibfield  {title} {\bibinfo {title} {Subgap states in
  disordered superconductors},\ }\href
  {https://doi.org/10.1134/S106377611311006X} {\bibfield  {journal} {\bibinfo
  {journal} {J. Exp. Theor. Phys.}\ }\textbf {\bibinfo {volume} {117}},\
  \bibinfo {pages} {487} (\bibinfo {year} {2013})}\BibitemShut {NoStop}%
\bibitem [{\citenamefont {Fominov}\ and\ \citenamefont
  {Skvortsov}(2016)}]{Fominov.2016}%
  \BibitemOpen
  \bibfield  {author} {\bibinfo {author} {\bibfnamefont {Y.~V.}\ \bibnamefont
  {Fominov}}\ and\ \bibinfo {author} {\bibfnamefont {M.~A.}\ \bibnamefont
  {Skvortsov}},\ }\bibfield  {title} {\bibinfo {title} {Subgap states in
  disordered superconductors with strong magnetic impurities},\ }\href
  {https://doi.org/10.1103/PhysRevB.93.144511} {\bibfield  {journal} {\bibinfo
  {journal} {Phys. Rev. B}\ }\textbf {\bibinfo {volume} {93}},\ \bibinfo
  {pages} {144511} (\bibinfo {year} {2016})}\BibitemShut {NoStop}%
\bibitem [{sup()}]{supplemental}%
  \BibitemOpen
  \href@noop {} {\bibinfo {title} {See {Supplemental Material} at [{URL} will
  be inserted by publisher], where
  {Refs.}~\cite{Giazotto.2006,Golubov.1993,Savich.2017,Zittartz.1966} are
  cited, for details on the generalized {RT} model and the subgap density of
  states.}}\BibitemShut {Stop}%
\bibitem [{\citenamefont {de~Rooij}\ \emph {et~al.}(2024)\citenamefont
  {de~Rooij}, \citenamefont {Fermin}, \citenamefont {Kouwenhoven},
  \citenamefont {Coppens}, \citenamefont {Murugesan}, \citenamefont {Thoen},
  \citenamefont {Aarts}, \citenamefont {Baselmans},\ and\ \citenamefont
  {de~Visser}}]{deRooij.2024}%
  \BibitemOpen
  \bibfield  {author} {\bibinfo {author} {\bibfnamefont {S.~A.~H.}\
  \bibnamefont {de~Rooij}}, \bibinfo {author} {\bibfnamefont {R.}~\bibnamefont
  {Fermin}}, \bibinfo {author} {\bibfnamefont {K.}~\bibnamefont {Kouwenhoven}},
  \bibinfo {author} {\bibfnamefont {T.}~\bibnamefont {Coppens}}, \bibinfo
  {author} {\bibfnamefont {V.}~\bibnamefont {Murugesan}}, \bibinfo {author}
  {\bibfnamefont {D.~J.}\ \bibnamefont {Thoen}}, \bibinfo {author}
  {\bibfnamefont {J.}~\bibnamefont {Aarts}}, \bibinfo {author} {\bibfnamefont
  {J.~J.~A.}\ \bibnamefont {Baselmans}},\ and\ \bibinfo {author} {\bibfnamefont
  {P.~J.}\ \bibnamefont {de~Visser}},\ }\bibfield  {title} {\bibinfo {title}
  {Recombination of localized quasiparticles in disordered superconductors},\
  }\href {https://arxiv.org/abs/2410.18802} {\bibfield  {journal} {\bibinfo
  {journal} {arXiv:2410.18802}\ } (\bibinfo {year} {2024})}\BibitemShut
  {NoStop}%
\bibitem [{\citenamefont {Kaplan}\ \emph {et~al.}(1976)\citenamefont {Kaplan},
  \citenamefont {Chi}, \citenamefont {Langenberg}, \citenamefont {Chang},
  \citenamefont {Jafarey},\ and\ \citenamefont {Scalapino}}]{Kaplan.1976}%
  \BibitemOpen
  \bibfield  {author} {\bibinfo {author} {\bibfnamefont {S.~B.}\ \bibnamefont
  {Kaplan}}, \bibinfo {author} {\bibfnamefont {C.~C.}\ \bibnamefont {Chi}},
  \bibinfo {author} {\bibfnamefont {D.~N.}\ \bibnamefont {Langenberg}},
  \bibinfo {author} {\bibfnamefont {J.~J.}\ \bibnamefont {Chang}}, \bibinfo
  {author} {\bibfnamefont {S.}~\bibnamefont {Jafarey}},\ and\ \bibinfo {author}
  {\bibfnamefont {D.~J.}\ \bibnamefont {Scalapino}},\ }\bibfield  {title}
  {\bibinfo {title} {Quasiparticle and phonon lifetimes in superconductors},\
  }\href {https://doi.org/10.1103/PhysRevB.14.4854} {\bibfield  {journal}
  {\bibinfo  {journal} {Phys. Rev. B}\ }\textbf {\bibinfo {volume} {14}},\
  \bibinfo {pages} {4854} (\bibinfo {year} {1976})}\BibitemShut {NoStop}%
\bibitem [{\citenamefont {{G. Catelani}}\ and\ \citenamefont {{D. M.
  Basko}}(2019)}]{G.Catelani.2019}%
  \BibitemOpen
  \bibfield  {author} {\bibinfo {author} {\bibnamefont {{G. Catelani}}}\ and\
  \bibinfo {author} {\bibnamefont {{D. M. Basko}}},\ }\bibfield  {title}
  {\bibinfo {title} {Non-equilibrium quasiparticles in superconducting
  circuits: photons vs. phonons},\ }\href
  {https://doi.org/10.21468/SciPostPhys.6.1.013} {\bibfield  {journal}
  {\bibinfo  {journal} {SciPost Phys}\ }\textbf {\bibinfo {volume} {6}},\
  \bibinfo {pages} {13} (\bibinfo {year} {2019})}\BibitemShut {NoStop}%
\bibitem [{\citenamefont {Fischer}\ and\ \citenamefont
  {Catelani}(2024)}]{Fischer.2024}%
  \BibitemOpen
  \bibfield  {author} {\bibinfo {author} {\bibfnamefont {P.~B.}\ \bibnamefont
  {Fischer}}\ and\ \bibinfo {author} {\bibfnamefont {G.}~\bibnamefont
  {Catelani}},\ }\bibfield  {title} {\bibinfo {title} {{Nonequilibrium
  quasiparticle distribution in superconducting resonators: Effect of
  pair-breaking photons}},\ }\href
  {https://doi.org/10.21468/SciPostPhys.17.3.070} {\bibfield  {journal}
  {\bibinfo  {journal} {SciPost Phys.}\ }\textbf {\bibinfo {volume} {17}},\
  \bibinfo {pages} {070} (\bibinfo {year} {2024})}\BibitemShut {NoStop}%
\bibitem [{\citenamefont {Larkin}\ and\ \citenamefont
  {Ovchinnikov}(1986)}]{Larkin.1986}%
  \BibitemOpen
  \bibfield  {author} {\bibinfo {author} {\bibfnamefont {A.~I.}\ \bibnamefont
  {Larkin}}\ and\ \bibinfo {author} {\bibfnamefont {Y.~N.}\ \bibnamefont
  {Ovchinnikov}},\ }\bibfield  {title} {\bibinfo {title} {Vortex motion in
  superconductors},\ }in\ \href@noop {} {\emph {\bibinfo {booktitle}
  {Nonequilibrium Superconductivity}}},\ \bibinfo {series and number} {Modern
  Problems in condensed matter science},\ \bibinfo {editor} {edited by\
  \bibinfo {editor} {\bibfnamefont {D.}~\bibnamefont {Langenberg}}\ and\
  \bibinfo {editor} {\bibfnamefont {A.~I.}\ \bibnamefont {Larkin}}}\ (\bibinfo
  {publisher} {North-Holland Publishing Company},\ \bibinfo {year}
  {1986})\BibitemShut {NoStop}%
\bibitem [{\citenamefont {Chang}\ and\ \citenamefont
  {Scalapino}(1978)}]{Chang.1978}%
  \BibitemOpen
  \bibfield  {author} {\bibinfo {author} {\bibfnamefont {J.}~\bibnamefont
  {Chang}}\ and\ \bibinfo {author} {\bibfnamefont {D.}~\bibnamefont
  {Scalapino}},\ }\bibfield  {title} {\bibinfo {title} {Nonequilibrium
  superconductivity},\ }\href {https://doi.org/10.1007/BF00116228} {\bibfield
  {journal} {\bibinfo  {journal} {J. Low Temp. Phys.}\ }\textbf {\bibinfo
  {volume} {31}},\ \bibinfo {pages} {1} (\bibinfo {year} {1978})}\BibitemShut
  {NoStop}%
\bibitem [{\citenamefont {Barends}\ \emph {et~al.}(2009)\citenamefont
  {Barends}, \citenamefont {van Vliet}, \citenamefont {Baselmans},
  \citenamefont {Yates}, \citenamefont {Gao},\ and\ \citenamefont
  {Klapwijk}}]{Barends.2009}%
  \BibitemOpen
  \bibfield  {author} {\bibinfo {author} {\bibfnamefont {R.}~\bibnamefont
  {Barends}}, \bibinfo {author} {\bibfnamefont {S.}~\bibnamefont {van Vliet}},
  \bibinfo {author} {\bibfnamefont {J.~J.~A.}\ \bibnamefont {Baselmans}},
  \bibinfo {author} {\bibfnamefont {S.~J.~C.}\ \bibnamefont {Yates}}, \bibinfo
  {author} {\bibfnamefont {J.~R.}\ \bibnamefont {Gao}},\ and\ \bibinfo {author}
  {\bibfnamefont {T.~M.}\ \bibnamefont {Klapwijk}},\ }\bibfield  {title}
  {\bibinfo {title} {Enhancement of quasiparticle recombination in ta and al
  superconductors by implantation of magnetic and nonmagnetic atoms},\ }\href
  {https://doi.org/10.1103/PhysRevB.79.020509} {\bibfield  {journal} {\bibinfo
  {journal} {Phys. Rev. B}\ }\textbf {\bibinfo {volume} {79}},\ \bibinfo
  {pages} {020509} (\bibinfo {year} {2009})}\BibitemShut {NoStop}%
\bibitem [{\citenamefont {Riwar}\ \emph {et~al.}(2016)\citenamefont {Riwar},
  \citenamefont {Hosseinkhani}, \citenamefont {Burkhart}, \citenamefont {Gao},
  \citenamefont {Schoelkopf}, \citenamefont {Glazman},\ and\ \citenamefont
  {Catelani}}]{Riwar.2016}%
  \BibitemOpen
  \bibfield  {author} {\bibinfo {author} {\bibfnamefont {R.-P.}\ \bibnamefont
  {Riwar}}, \bibinfo {author} {\bibfnamefont {A.}~\bibnamefont {Hosseinkhani}},
  \bibinfo {author} {\bibfnamefont {L.~D.}\ \bibnamefont {Burkhart}}, \bibinfo
  {author} {\bibfnamefont {Y.~Y.}\ \bibnamefont {Gao}}, \bibinfo {author}
  {\bibfnamefont {R.~J.}\ \bibnamefont {Schoelkopf}}, \bibinfo {author}
  {\bibfnamefont {L.~I.}\ \bibnamefont {Glazman}},\ and\ \bibinfo {author}
  {\bibfnamefont {G.}~\bibnamefont {Catelani}},\ }\bibfield  {title} {\bibinfo
  {title} {Normal-metal quasiparticle traps for superconducting qubits},\
  }\href {https://doi.org/10.1103/PhysRevB.94.104516} {\bibfield  {journal}
  {\bibinfo  {journal} {Phys. Rev. B}\ }\textbf {\bibinfo {volume} {94}},\
  \bibinfo {pages} {104516} (\bibinfo {year} {2016})}\BibitemShut {NoStop}%
\bibitem [{\citenamefont {Fischer}\ and\ \citenamefont
  {Catelani}(2023)}]{Fischer.2023}%
  \BibitemOpen
  \bibfield  {author} {\bibinfo {author} {\bibfnamefont {P.}~\bibnamefont
  {Fischer}}\ and\ \bibinfo {author} {\bibfnamefont {G.}~\bibnamefont
  {Catelani}},\ }\bibfield  {title} {\bibinfo {title} {Nonequilibrium
  quasiparticle distribution in superconducting resonators: An analytical
  approach},\ }\href {https://doi.org/10.1103/PhysRevApplied.19.054087}
  {\bibfield  {journal} {\bibinfo  {journal} {Phys. Rev. Appl.}\ }\textbf
  {\bibinfo {volume} {19}},\ \bibinfo {pages} {054087} (\bibinfo {year}
  {2023})}\BibitemShut {NoStop}%
\bibitem [{\citenamefont {Wang}\ \emph {et~al.}(2014)\citenamefont {Wang},
  \citenamefont {Gao}, \citenamefont {Pop}, \citenamefont {Vool}, \citenamefont
  {Axline}, \citenamefont {Brecht}, \citenamefont {Heeres}, \citenamefont
  {Frunzio}, \citenamefont {Devoret}, \citenamefont {Catelani}, \citenamefont
  {Glazman},\ and\ \citenamefont {Schoelkopf}}]{Wang.2014}%
  \BibitemOpen
  \bibfield  {author} {\bibinfo {author} {\bibfnamefont {C.}~\bibnamefont
  {Wang}}, \bibinfo {author} {\bibfnamefont {Y.~Y.}\ \bibnamefont {Gao}},
  \bibinfo {author} {\bibfnamefont {I.~M.}\ \bibnamefont {Pop}}, \bibinfo
  {author} {\bibfnamefont {U.}~\bibnamefont {Vool}}, \bibinfo {author}
  {\bibfnamefont {C.}~\bibnamefont {Axline}}, \bibinfo {author} {\bibfnamefont
  {T.}~\bibnamefont {Brecht}}, \bibinfo {author} {\bibfnamefont {R.~W.}\
  \bibnamefont {Heeres}}, \bibinfo {author} {\bibfnamefont {L.}~\bibnamefont
  {Frunzio}}, \bibinfo {author} {\bibfnamefont {M.~H.}\ \bibnamefont
  {Devoret}}, \bibinfo {author} {\bibfnamefont {G.}~\bibnamefont {Catelani}},
  \bibinfo {author} {\bibfnamefont {L.~I.}\ \bibnamefont {Glazman}},\ and\
  \bibinfo {author} {\bibfnamefont {R.~J.}\ \bibnamefont {Schoelkopf}},\
  }\bibfield  {title} {\bibinfo {title} {Measurement and control of
  quasiparticle dynamics in a superconducting qubit},\ }\href
  {https://doi.org/10.1038/ncomms6836} {\bibfield  {journal} {\bibinfo
  {journal} {Nat. Commun.}\ }\textbf {\bibinfo {volume} {5}},\ \bibinfo {pages}
  {5836} (\bibinfo {year} {2014})}\BibitemShut {NoStop}%
\bibitem [{\citenamefont {Hosseinkhani}\ and\ \citenamefont
  {Catelani}(2018)}]{Hosseinkhani.2018}%
  \BibitemOpen
  \bibfield  {author} {\bibinfo {author} {\bibfnamefont {A.}~\bibnamefont
  {Hosseinkhani}}\ and\ \bibinfo {author} {\bibfnamefont {G.}~\bibnamefont
  {Catelani}},\ }\bibfield  {title} {\bibinfo {title} {Proximity effect in
  normal-metal quasiparticle traps},\ }\href
  {https://doi.org/10.1103/PhysRevB.97.054513} {\bibfield  {journal} {\bibinfo
  {journal} {Phys. Rev. B}\ }\textbf {\bibinfo {volume} {97}},\ \bibinfo
  {pages} {054513} (\bibinfo {year} {2018})}\BibitemShut {NoStop}%
\bibitem [{\citenamefont {O’Neil}\ \emph {et~al.}(2010)\citenamefont
  {O’Neil}, \citenamefont {Schmidt}, \citenamefont {Tomlin},\ and\
  \citenamefont {Ullom}}]{ONeil.2010}%
  \BibitemOpen
  \bibfield  {author} {\bibinfo {author} {\bibfnamefont {G.~C.}\ \bibnamefont
  {O’Neil}}, \bibinfo {author} {\bibfnamefont {D.~R.}\ \bibnamefont
  {Schmidt}}, \bibinfo {author} {\bibfnamefont {N.~A.}\ \bibnamefont
  {Tomlin}},\ and\ \bibinfo {author} {\bibfnamefont {J.~N.}\ \bibnamefont
  {Ullom}},\ }\bibfield  {title} {\bibinfo {title} {Quasiparticle density of
  states measurements in clean superconducting {AlMn} alloys},\ }\href
  {https://doi.org/10.1063/1.3369280} {\bibfield  {journal} {\bibinfo
  {journal} {J. Appl. Phys.}\ }\textbf {\bibinfo {volume} {107}},\ \bibinfo
  {pages} {093903} (\bibinfo {year} {2010})}\BibitemShut {NoStop}%
\bibitem [{\citenamefont {Fornieri}\ \emph {et~al.}(2015)\citenamefont
  {Fornieri}, \citenamefont {Blanc}, \citenamefont {Bosisio}, \citenamefont
  {D'Ambrosio},\ and\ \citenamefont {Giazotto}}]{NNano.2015}%
  \BibitemOpen
  \bibfield  {author} {\bibinfo {author} {\bibfnamefont {A.}~\bibnamefont
  {Fornieri}}, \bibinfo {author} {\bibfnamefont {C.}~\bibnamefont {Blanc}},
  \bibinfo {author} {\bibfnamefont {R.}~\bibnamefont {Bosisio}}, \bibinfo
  {author} {\bibfnamefont {S.}~\bibnamefont {D'Ambrosio}},\ and\ \bibinfo
  {author} {\bibfnamefont {F.}~\bibnamefont {Giazotto}},\ }\bibfield  {title}
  {\bibinfo {title} {Nanoscale phase engineering of thermal transport with a
  {Josephson} heat modulator},\ }\href {https://doi.org/10.1038/nnano.2015.281}
  {\bibfield  {journal} {\bibinfo  {journal} {Nat. Nanotechnol.}\ }\textbf
  {\bibinfo {volume} {11}},\ \bibinfo {pages} {258} (\bibinfo {year}
  {2015})}\BibitemShut {NoStop}%
\bibitem [{\citenamefont {Fornieri}\ \emph {et~al.}(2017)\citenamefont
  {Fornieri}, \citenamefont {Timossi}, \citenamefont {Virtanen}, \citenamefont
  {Solinas},\ and\ \citenamefont {Giazotto}}]{NNano.2017}%
  \BibitemOpen
  \bibfield  {author} {\bibinfo {author} {\bibfnamefont {A.}~\bibnamefont
  {Fornieri}}, \bibinfo {author} {\bibfnamefont {G.}~\bibnamefont {Timossi}},
  \bibinfo {author} {\bibfnamefont {P.}~\bibnamefont {Virtanen}}, \bibinfo
  {author} {\bibfnamefont {P.}~\bibnamefont {Solinas}},\ and\ \bibinfo {author}
  {\bibfnamefont {F.}~\bibnamefont {Giazotto}},\ }\bibfield  {title} {\bibinfo
  {title} {$0$–$\pi$ phase-controllable thermal {Josephson} junction},\
  }\href {https://doi.org/10.1038/nnano.2017.25} {\bibfield  {journal}
  {\bibinfo  {journal} {Nat. Nanotechnol.}\ }\textbf {\bibinfo {volume} {12}},\
  \bibinfo {pages} {425} (\bibinfo {year} {2017})}\BibitemShut {NoStop}%
\bibitem [{\citenamefont {Germanese}\ \emph {et~al.}(2022)\citenamefont
  {Germanese}, \citenamefont {Paolucci}, \citenamefont {Marchegiani},
  \citenamefont {Braggio},\ and\ \citenamefont {Giazotto}}]{NNano.2022}%
  \BibitemOpen
  \bibfield  {author} {\bibinfo {author} {\bibfnamefont {G.}~\bibnamefont
  {Germanese}}, \bibinfo {author} {\bibfnamefont {F.}~\bibnamefont {Paolucci}},
  \bibinfo {author} {\bibfnamefont {G.}~\bibnamefont {Marchegiani}}, \bibinfo
  {author} {\bibfnamefont {A.}~\bibnamefont {Braggio}},\ and\ \bibinfo {author}
  {\bibfnamefont {F.}~\bibnamefont {Giazotto}},\ }\bibfield  {title} {\bibinfo
  {title} {Bipolar thermoelectric {Josephson} engine},\ }\href
  {https://doi.org/10.1038/s41565-022-01208-y} {\bibfield  {journal} {\bibinfo
  {journal} {Nat. Nanotechnol.}\ }\textbf {\bibinfo {volume} {17}},\ \bibinfo
  {pages} {1084} (\bibinfo {year} {2022})}\BibitemShut {NoStop}%
\bibitem [{\citenamefont {Paik}\ \emph {et~al.}(2011)\citenamefont {Paik},
  \citenamefont {Schuster}, \citenamefont {Bishop}, \citenamefont {Kirchmair},
  \citenamefont {Catelani}, \citenamefont {Sears}, \citenamefont {Johnson},
  \citenamefont {Reagor}, \citenamefont {Frunzio}, \citenamefont {Glazman},
  \citenamefont {Girvin}, \citenamefont {Devoret},\ and\ \citenamefont
  {Schoelkopf}}]{Paik.2011}%
  \BibitemOpen
  \bibfield  {author} {\bibinfo {author} {\bibfnamefont {H.}~\bibnamefont
  {Paik}}, \bibinfo {author} {\bibfnamefont {D.~I.}\ \bibnamefont {Schuster}},
  \bibinfo {author} {\bibfnamefont {L.~S.}\ \bibnamefont {Bishop}}, \bibinfo
  {author} {\bibfnamefont {G.}~\bibnamefont {Kirchmair}}, \bibinfo {author}
  {\bibfnamefont {G.}~\bibnamefont {Catelani}}, \bibinfo {author}
  {\bibfnamefont {A.~P.}\ \bibnamefont {Sears}}, \bibinfo {author}
  {\bibfnamefont {B.~R.}\ \bibnamefont {Johnson}}, \bibinfo {author}
  {\bibfnamefont {M.~J.}\ \bibnamefont {Reagor}}, \bibinfo {author}
  {\bibfnamefont {L.}~\bibnamefont {Frunzio}}, \bibinfo {author} {\bibfnamefont
  {L.~I.}\ \bibnamefont {Glazman}}, \bibinfo {author} {\bibfnamefont {S.~M.}\
  \bibnamefont {Girvin}}, \bibinfo {author} {\bibfnamefont {M.~H.}\
  \bibnamefont {Devoret}},\ and\ \bibinfo {author} {\bibfnamefont {R.~J.}\
  \bibnamefont {Schoelkopf}},\ }\bibfield  {title} {\bibinfo {title}
  {Observation of high coherence in {Josephson} junction qubits measured in a
  three-dimensional circuit {QED} architecture},\ }\href
  {https://doi.org/10.1103/PhysRevLett.107.240501} {\bibfield  {journal}
  {\bibinfo  {journal} {Phys. Rev. Lett.}\ }\textbf {\bibinfo {volume} {107}},\
  \bibinfo {pages} {240501} (\bibinfo {year} {2011})}\BibitemShut {NoStop}%
\bibitem [{\citenamefont {de~Visser}\ \emph {et~al.}(2014)\citenamefont
  {de~Visser}, \citenamefont {Goldie}, \citenamefont {Diener}, \citenamefont
  {Withington}, \citenamefont {Baselmans},\ and\ \citenamefont
  {Klapwijk}}]{Visser.2014}%
  \BibitemOpen
  \bibfield  {author} {\bibinfo {author} {\bibfnamefont {P.~J.}\ \bibnamefont
  {de~Visser}}, \bibinfo {author} {\bibfnamefont {D.~J.}\ \bibnamefont
  {Goldie}}, \bibinfo {author} {\bibfnamefont {P.}~\bibnamefont {Diener}},
  \bibinfo {author} {\bibfnamefont {S.}~\bibnamefont {Withington}}, \bibinfo
  {author} {\bibfnamefont {J.~J.~A.}\ \bibnamefont {Baselmans}},\ and\ \bibinfo
  {author} {\bibfnamefont {T.~M.}\ \bibnamefont {Klapwijk}},\ }\bibfield
  {title} {\bibinfo {title} {Evidence of a nonequilibrium distribution of
  quasiparticles in the microwave response of a superconducting aluminum
  resonator},\ }\href {https://doi.org/10.1103/PhysRevLett.112.047004}
  {\bibfield  {journal} {\bibinfo  {journal} {Phys. Rev. Lett.}\ }\textbf
  {\bibinfo {volume} {112}},\ \bibinfo {pages} {047004} (\bibinfo {year}
  {2014})}\BibitemShut {NoStop}%
\bibitem [{\citenamefont {Dynes}\ and\ \citenamefont
  {Garno}(1981)}]{Dynes.1981}%
  \BibitemOpen
  \bibfield  {author} {\bibinfo {author} {\bibfnamefont {R.~C.}\ \bibnamefont
  {Dynes}}\ and\ \bibinfo {author} {\bibfnamefont {J.~P.}\ \bibnamefont
  {Garno}},\ }\bibfield  {title} {\bibinfo {title} {Metal-insulator transition
  in granular aluminum},\ }\href {https://doi.org/10.1103/PhysRevLett.46.137}
  {\bibfield  {journal} {\bibinfo  {journal} {Phys. Rev. Lett.}\ }\textbf
  {\bibinfo {volume} {46}},\ \bibinfo {pages} {137} (\bibinfo {year}
  {1981})}\BibitemShut {NoStop}%
\bibitem [{\citenamefont {Valenti}\ \emph {et~al.}(2019)\citenamefont
  {Valenti}, \citenamefont {Henriques}, \citenamefont {Catelani}, \citenamefont
  {Maleeva}, \citenamefont {Gr\"unhaupt}, \citenamefont {von L\"upke},
  \citenamefont {Skacel}, \citenamefont {Winkel}, \citenamefont {Bilmes},
  \citenamefont {Ustinov}, \citenamefont {Goupy}, \citenamefont {Calvo},
  \citenamefont {Beno\^{\i}t}, \citenamefont {Levy-Bertrand}, \citenamefont
  {Monfardini},\ and\ \citenamefont {Pop}}]{Valenti.2019}%
  \BibitemOpen
  \bibfield  {author} {\bibinfo {author} {\bibfnamefont {F.}~\bibnamefont
  {Valenti}}, \bibinfo {author} {\bibfnamefont {F.}~\bibnamefont {Henriques}},
  \bibinfo {author} {\bibfnamefont {G.}~\bibnamefont {Catelani}}, \bibinfo
  {author} {\bibfnamefont {N.}~\bibnamefont {Maleeva}}, \bibinfo {author}
  {\bibfnamefont {L.}~\bibnamefont {Gr\"unhaupt}}, \bibinfo {author}
  {\bibfnamefont {U.}~\bibnamefont {von L\"upke}}, \bibinfo {author}
  {\bibfnamefont {S.~T.}\ \bibnamefont {Skacel}}, \bibinfo {author}
  {\bibfnamefont {P.}~\bibnamefont {Winkel}}, \bibinfo {author} {\bibfnamefont
  {A.}~\bibnamefont {Bilmes}}, \bibinfo {author} {\bibfnamefont {A.~V.}\
  \bibnamefont {Ustinov}}, \bibinfo {author} {\bibfnamefont {J.}~\bibnamefont
  {Goupy}}, \bibinfo {author} {\bibfnamefont {M.}~\bibnamefont {Calvo}},
  \bibinfo {author} {\bibfnamefont {A.}~\bibnamefont {Beno\^{\i}t}}, \bibinfo
  {author} {\bibfnamefont {F.}~\bibnamefont {Levy-Bertrand}}, \bibinfo {author}
  {\bibfnamefont {A.}~\bibnamefont {Monfardini}},\ and\ \bibinfo {author}
  {\bibfnamefont {I.~M.}\ \bibnamefont {Pop}},\ }\bibfield  {title} {\bibinfo
  {title} {Interplay between kinetic inductance, nonlinearity, and
  quasiparticle dynamics in granular aluminum microwave kinetic inductance
  detectors},\ }\href {https://doi.org/10.1103/PhysRevApplied.11.054087}
  {\bibfield  {journal} {\bibinfo  {journal} {Phys. Rev. Appl.}\ }\textbf
  {\bibinfo {volume} {11}},\ \bibinfo {pages} {054087} (\bibinfo {year}
  {2019})}\BibitemShut {NoStop}%
\bibitem [{\citenamefont {Dynes}\ \emph {et~al.}(1984)\citenamefont {Dynes},
  \citenamefont {Garno}, \citenamefont {Hertel},\ and\ \citenamefont
  {Orlando}}]{Dynes.1984}%
  \BibitemOpen
  \bibfield  {author} {\bibinfo {author} {\bibfnamefont {R.~C.}\ \bibnamefont
  {Dynes}}, \bibinfo {author} {\bibfnamefont {J.~P.}\ \bibnamefont {Garno}},
  \bibinfo {author} {\bibfnamefont {G.~B.}\ \bibnamefont {Hertel}},\ and\
  \bibinfo {author} {\bibfnamefont {T.~P.}\ \bibnamefont {Orlando}},\
  }\bibfield  {title} {\bibinfo {title} {Tunneling study of superconductivity
  near the metal-insulator transition},\ }\href
  {https://doi.org/10.1103/PhysRevLett.53.2437} {\bibfield  {journal} {\bibinfo
   {journal} {Phys. Rev. Lett.}\ }\textbf {\bibinfo {volume} {53}},\ \bibinfo
  {pages} {2437} (\bibinfo {year} {1984})}\BibitemShut {NoStop}%
\bibitem [{\citenamefont {Larson}\ \emph {et~al.}(2025)\citenamefont {Larson},
  \citenamefont {Jones}, \citenamefont {Kalmar}, \citenamefont {Sanchez},
  \citenamefont {Chitta}, \citenamefont {Verma}, \citenamefont {Genter},
  \citenamefont {Cicak}, \citenamefont {Nam}, \citenamefont {Fulop},
  \citenamefont {Koch}, \citenamefont {Simmonds},\ and\ \citenamefont
  {Gyenis}}]{Larson.2025}%
  \BibitemOpen
  \bibfield  {author} {\bibinfo {author} {\bibfnamefont {T.~F.~Q.}\
  \bibnamefont {Larson}}, \bibinfo {author} {\bibfnamefont {S.~G.}\
  \bibnamefont {Jones}}, \bibinfo {author} {\bibfnamefont {T.}~\bibnamefont
  {Kalmar}}, \bibinfo {author} {\bibfnamefont {P.~A.}\ \bibnamefont {Sanchez}},
  \bibinfo {author} {\bibfnamefont {S.~P.}\ \bibnamefont {Chitta}}, \bibinfo
  {author} {\bibfnamefont {V.}~\bibnamefont {Verma}}, \bibinfo {author}
  {\bibfnamefont {K.}~\bibnamefont {Genter}}, \bibinfo {author} {\bibfnamefont
  {K.}~\bibnamefont {Cicak}}, \bibinfo {author} {\bibfnamefont {S.~W.}\
  \bibnamefont {Nam}}, \bibinfo {author} {\bibfnamefont {G.}~\bibnamefont
  {Fulop}}, \bibinfo {author} {\bibfnamefont {J.}~\bibnamefont {Koch}},
  \bibinfo {author} {\bibfnamefont {R.~W.}\ \bibnamefont {Simmonds}},\ and\
  \bibinfo {author} {\bibfnamefont {A.}~\bibnamefont {Gyenis}},\ }\bibfield
  {title} {\bibinfo {title} {Localized quasiparticles in a fluxonium with
  quasi-two-dimensional amorphous kinetic inductors},\ }\href
  {https://arxiv.org/abs/2504.07950} {\bibfield  {journal} {\bibinfo  {journal}
  {arXiv:2504.07950}\ } (\bibinfo {year} {2025})}\BibitemShut {NoStop}%
\bibitem [{\citenamefont {Nevala}\ \emph {et~al.}(2012)\citenamefont {Nevala},
  \citenamefont {Chaudhuri}, \citenamefont {Halkosaari}, \citenamefont
  {Karvonen},\ and\ \citenamefont {Maasilta}}]{Nevala.2012}%
  \BibitemOpen
  \bibfield  {author} {\bibinfo {author} {\bibfnamefont {M.~R.}\ \bibnamefont
  {Nevala}}, \bibinfo {author} {\bibfnamefont {S.}~\bibnamefont {Chaudhuri}},
  \bibinfo {author} {\bibfnamefont {J.}~\bibnamefont {Halkosaari}}, \bibinfo
  {author} {\bibfnamefont {J.~T.}\ \bibnamefont {Karvonen}},\ and\ \bibinfo
  {author} {\bibfnamefont {I.~J.}\ \bibnamefont {Maasilta}},\ }\bibfield
  {title} {\bibinfo {title} {Sub-micron normal-metal/insulator/superconductor
  tunnel junction thermometer and cooler using {Nb}},\ }\href
  {https://doi.org/10.1063/1.4751355} {\bibfield  {journal} {\bibinfo
  {journal} {Applied Physics Letters}\ }\textbf {\bibinfo {volume} {101}},\
  \bibinfo {pages} {112601} (\bibinfo {year} {2012})}\BibitemShut {NoStop}%
\bibitem [{\citenamefont {Chaudhuri}\ \emph {et~al.}(2013)\citenamefont
  {Chaudhuri}, \citenamefont {Nevala},\ and\ \citenamefont
  {Maasilta}}]{Chaudhuri.2013}%
  \BibitemOpen
  \bibfield  {author} {\bibinfo {author} {\bibfnamefont {S.}~\bibnamefont
  {Chaudhuri}}, \bibinfo {author} {\bibfnamefont {M.~R.}\ \bibnamefont
  {Nevala}},\ and\ \bibinfo {author} {\bibfnamefont {I.~J.}\ \bibnamefont
  {Maasilta}},\ }\bibfield  {title} {\bibinfo {title} {Niobium nitride-based
  normal metal-insulator-superconductor tunnel junction microthermometer},\
  }\href {https://doi.org/10.1063/1.4800440} {\bibfield  {journal} {\bibinfo
  {journal} {Applied Physics Letters}\ }\textbf {\bibinfo {volume} {102}},\
  \bibinfo {pages} {132601} (\bibinfo {year} {2013})}\BibitemShut {NoStop}%
\bibitem [{\citenamefont {Chaudhuri}\ and\ \citenamefont
  {Maasilta}(2014)}]{Chaudhuri.2014}%
  \BibitemOpen
  \bibfield  {author} {\bibinfo {author} {\bibfnamefont {S.}~\bibnamefont
  {Chaudhuri}}\ and\ \bibinfo {author} {\bibfnamefont {I.~J.}\ \bibnamefont
  {Maasilta}},\ }\bibfield  {title} {\bibinfo {title} {Superconducting tantalum
  nitride-based normal metal-insulator-superconductor tunnel junctions},\
  }\href {https://doi.org/10.1063/1.4869563} {\bibfield  {journal} {\bibinfo
  {journal} {Applied Physics Letters}\ }\textbf {\bibinfo {volume} {104}},\
  \bibinfo {pages} {122601} (\bibinfo {year} {2014})}\BibitemShut {NoStop}%
\bibitem [{\citenamefont {de~Graaf}\ \emph {et~al.}(2020)\citenamefont
  {de~Graaf}, \citenamefont {Faoro}, \citenamefont {Ioffe}, \citenamefont
  {Mahashabde}, \citenamefont {Burnett}, \citenamefont {Lindstr{\"o}m},
  \citenamefont {Kubatkin}, \citenamefont {Danilov},\ and\ \citenamefont
  {Tzalenchuk}}]{Graaf.2020}%
  \BibitemOpen
  \bibfield  {author} {\bibinfo {author} {\bibfnamefont {S.~E.}\ \bibnamefont
  {de~Graaf}}, \bibinfo {author} {\bibfnamefont {L.}~\bibnamefont {Faoro}},
  \bibinfo {author} {\bibfnamefont {L.~B.}\ \bibnamefont {Ioffe}}, \bibinfo
  {author} {\bibfnamefont {S.}~\bibnamefont {Mahashabde}}, \bibinfo {author}
  {\bibfnamefont {J.~J.}\ \bibnamefont {Burnett}}, \bibinfo {author}
  {\bibfnamefont {T.}~\bibnamefont {Lindstr{\"o}m}}, \bibinfo {author}
  {\bibfnamefont {S.~E.}\ \bibnamefont {Kubatkin}}, \bibinfo {author}
  {\bibfnamefont {A.~V.}\ \bibnamefont {Danilov}},\ and\ \bibinfo {author}
  {\bibfnamefont {A.~Y.}\ \bibnamefont {Tzalenchuk}},\ }\bibfield  {title}
  {\bibinfo {title} {Two-level systems in superconducting quantum devices due
  to trapped quasiparticles},\ }\href {https://doi.org/10.1126/sciadv.abc5055}
  {\bibfield  {journal} {\bibinfo  {journal} {Sci. Adv.}\ }\textbf {\bibinfo
  {volume} {6}},\ \bibinfo {pages} {eabc5055} (\bibinfo {year}
  {2020})}\BibitemShut {NoStop}%
\bibitem [{\citenamefont {Kozorezov}\ \emph {et~al.}(2008)\citenamefont
  {Kozorezov}, \citenamefont {Golubov}, \citenamefont {Wigmore}, \citenamefont
  {Martin}, \citenamefont {Verhoeve}, \citenamefont {Hijmering},\ and\
  \citenamefont {Jerjen}}]{Kozorezov.2008}%
  \BibitemOpen
  \bibfield  {author} {\bibinfo {author} {\bibfnamefont {A.~G.}\ \bibnamefont
  {Kozorezov}}, \bibinfo {author} {\bibfnamefont {A.~A.}\ \bibnamefont
  {Golubov}}, \bibinfo {author} {\bibfnamefont {J.~K.}\ \bibnamefont
  {Wigmore}}, \bibinfo {author} {\bibfnamefont {D.}~\bibnamefont {Martin}},
  \bibinfo {author} {\bibfnamefont {P.}~\bibnamefont {Verhoeve}}, \bibinfo
  {author} {\bibfnamefont {R.~A.}\ \bibnamefont {Hijmering}},\ and\ \bibinfo
  {author} {\bibfnamefont {I.}~\bibnamefont {Jerjen}},\ }\bibfield  {title}
  {\bibinfo {title} {Inelastic scattering of quasiparticles in a superconductor
  with magnetic impurities},\ }\href
  {https://doi.org/10.1103/PhysRevB.78.174501} {\bibfield  {journal} {\bibinfo
  {journal} {Phys. Rev. B}\ }\textbf {\bibinfo {volume} {78}},\ \bibinfo
  {pages} {174501} (\bibinfo {year} {2008})}\BibitemShut {NoStop}%
\bibitem [{\citenamefont {Hijmering}\ \emph {et~al.}(2009)\citenamefont
  {Hijmering}, \citenamefont {Kozorezov}, \citenamefont {Golubov},
  \citenamefont {Verhoeve}, \citenamefont {Martin}, \citenamefont {Wigmore},\
  and\ \citenamefont {Jerjen}}]{Hijmering.2009}%
  \BibitemOpen
  \bibfield  {author} {\bibinfo {author} {\bibfnamefont {R.~A.}\ \bibnamefont
  {Hijmering}}, \bibinfo {author} {\bibfnamefont {A.~G.}\ \bibnamefont
  {Kozorezov}}, \bibinfo {author} {\bibfnamefont {A.~A.}\ \bibnamefont
  {Golubov}}, \bibinfo {author} {\bibfnamefont {P.}~\bibnamefont {Verhoeve}},
  \bibinfo {author} {\bibfnamefont {D.~D.~E.}\ \bibnamefont {Martin}}, \bibinfo
  {author} {\bibfnamefont {J.~K.}\ \bibnamefont {Wigmore}},\ and\ \bibinfo
  {author} {\bibfnamefont {I.}~\bibnamefont {Jerjen}},\ }\bibfield  {title}
  {\bibinfo {title} {Modeling of local trapping states in superconductor tunnel
  junctions with kondo impurities},\ }\href
  {https://doi.org/10.1109/TASC.2009.2018506} {\bibfield  {journal} {\bibinfo
  {journal} {IEEE Trans. Appl. Supercond.}\ }\textbf {\bibinfo {volume} {19}},\
  \bibinfo {pages} {423} (\bibinfo {year} {2009})}\BibitemShut {NoStop}%
\bibitem [{\citenamefont {Gurra}\ \emph {et~al.}(2025)\citenamefont {Gurra},
  \citenamefont {Bennett}, \citenamefont {Duff}, \citenamefont {Vissers},\ and\
  \citenamefont {Ullom}}]{Ullom}%
  \BibitemOpen
  \bibfield  {author} {\bibinfo {author} {\bibfnamefont {E.}~\bibnamefont
  {Gurra}}, \bibinfo {author} {\bibfnamefont {D.~A.}\ \bibnamefont {Bennett}},
  \bibinfo {author} {\bibfnamefont {S.~M.}\ \bibnamefont {Duff}}, \bibinfo
  {author} {\bibfnamefont {M.~R.}\ \bibnamefont {Vissers}},\ and\ \bibinfo
  {author} {\bibfnamefont {J.~N.}\ \bibnamefont {Ullom}},\ }\bibfield  {title}
  {\bibinfo {title} {Can slow recombination in ordered superconductors explain
  the excess quasiparticle population?},\ }\href
  {https://arxiv.org/abs/2507.03217} {\bibfield  {journal} {\bibinfo  {journal}
  {arXiv:2507.03217}\ } (\bibinfo {year} {2025})}\BibitemShut {NoStop}%
\bibitem [{\citenamefont {Giazotto}\ \emph {et~al.}(2006)\citenamefont
  {Giazotto}, \citenamefont {Heikkil{\"a}}, \citenamefont {Luukanen},
  \citenamefont {Savin},\ and\ \citenamefont {Pekola}}]{Giazotto.2006}%
  \BibitemOpen
  \bibfield  {author} {\bibinfo {author} {\bibfnamefont {F.}~\bibnamefont
  {Giazotto}}, \bibinfo {author} {\bibfnamefont {T.~T.}\ \bibnamefont
  {Heikkil{\"a}}}, \bibinfo {author} {\bibfnamefont {A.}~\bibnamefont
  {Luukanen}}, \bibinfo {author} {\bibfnamefont {A.~M.}\ \bibnamefont
  {Savin}},\ and\ \bibinfo {author} {\bibfnamefont {J.~P.}\ \bibnamefont
  {Pekola}},\ }\bibfield  {title} {\bibinfo {title} {Opportunities for
  mesoscopics in thermometry and refrigeration: Physics and applications},\
  }\href {https://doi.org/10.1103/RevModPhys.78.217} {\bibfield  {journal}
  {\bibinfo  {journal} {Reviews of Modern Physics}\ }\textbf {\bibinfo {volume}
  {78}},\ \bibinfo {pages} {217} (\bibinfo {year} {2006})}\BibitemShut
  {NoStop}%
\bibitem [{\citenamefont {Golubov}\ and\ \citenamefont
  {Houwman}(1993)}]{Golubov.1993}%
  \BibitemOpen
  \bibfield  {author} {\bibinfo {author} {\bibfnamefont {A.}~\bibnamefont
  {Golubov}}\ and\ \bibinfo {author} {\bibfnamefont {E.}~\bibnamefont
  {Houwman}},\ }\bibfield  {title} {\bibinfo {title} {Quasiparticle relaxation
  rates in a spatially inhomogeneous superconductor},\ }\href
  {https://doi.org/https://doi.org/10.1016/0921-4534(93)90181-O} {\bibfield
  {journal} {\bibinfo  {journal} {Physica C: Superconductivity}\ }\textbf
  {\bibinfo {volume} {205}},\ \bibinfo {pages} {147} (\bibinfo {year}
  {1993})}\BibitemShut {NoStop}%
\bibitem [{\citenamefont {Savich}\ \emph {et~al.}(2017)\citenamefont {Savich},
  \citenamefont {Glazman},\ and\ \citenamefont {Kamenev}}]{Savich.2017}%
  \BibitemOpen
  \bibfield  {author} {\bibinfo {author} {\bibfnamefont {Y.}~\bibnamefont
  {Savich}}, \bibinfo {author} {\bibfnamefont {L.}~\bibnamefont {Glazman}},\
  and\ \bibinfo {author} {\bibfnamefont {A.}~\bibnamefont {Kamenev}},\
  }\bibfield  {title} {\bibinfo {title} {Quasiparticle relaxation in
  superconducting nanostructures},\ }\href
  {https://doi.org/10.1103/PhysRevB.96.104510} {\bibfield  {journal} {\bibinfo
  {journal} {Phys. Rev. B}\ }\textbf {\bibinfo {volume} {96}},\ \bibinfo
  {pages} {104510} (\bibinfo {year} {2017})}\BibitemShut {NoStop}%
\bibitem [{\citenamefont {Zittartz}\ and\ \citenamefont
  {Langer}(1966)}]{Zittartz.1966}%
  \BibitemOpen
  \bibfield  {author} {\bibinfo {author} {\bibfnamefont {J.}~\bibnamefont
  {Zittartz}}\ and\ \bibinfo {author} {\bibfnamefont {J.~S.}\ \bibnamefont
  {Langer}},\ }\bibfield  {title} {\bibinfo {title} {Theory of bound states in
  a random potential},\ }\href {https://doi.org/10.1103/PhysRev.148.741}
  {\bibfield  {journal} {\bibinfo  {journal} {Phys. Rev.}\ }\textbf {\bibinfo
  {volume} {148}},\ \bibinfo {pages} {741} (\bibinfo {year}
  {1966})}\BibitemShut {NoStop}%
\end{thebibliography}%

\onecolumngrid

\clearpage

\setcounter{figure}{0}

\makeatletter 
\renewcommand{\thefigure}{S\@arabic\c@figure}
\makeatother

\setcounter{page}{1}

\begin{center}

\textbf{\large{Supplemental Material for \\
``Excess quasiparticles and their dynamics in the presence of subgap states''}}

\vspace{0.4cm}

P. B. Fischer$^{1,2}$ and G. Catelani$^{1,3}$

\vspace{0.15cm}

$^1$\textit{\small{JARA Institute for Quantum Information (PGI-11),Forschungszentrum J\"ulich, 52425 J\"ulich, Germany}}

$^2$\textit{\small{JARA Institute for Quantum Information, RWTH Aachen University, 52056 Aachen, Germany}}

$^3$\textit{\small{Quantum Research Center, Technology Innovation Institute, Abu Dhabi 9639, UAE}}

\vspace{0.4cm}

\end{center}

\twocolumngrid

\appendix

\section{Generalized RT model}
\label{app:genRT}

The structure of the generalized RT model, Eqs.~\eqref{eq: Effective Rate Equation} and \eqref{eq:xm rate eq}, and the estimates for the rates appearing in it can be justified from microscopic theory. In a dirty superconductor with strong non-magnetic impurity scattering (mean-free path short compared to coherence length $\ell \ll \xi_0=v_F/\Delta_0$, with $v_F$ the Fermi velocity and $\Delta_0$ the zero-temperature superconducting gap) in the absence of electromagnetic fields and branch imbalance, the kinetic equation for the quasiparticle distribution function $f(E,t)$ has the form~\cite{Larkin.1986}
\begin{align}\label{eq: Diffusion equation}
   \rho(E) \frac{\partial f (E,t)}{\partial t}-D\vec{\nabla}\cdot\left(\rho(E)^2 K^{-}(E,E)\vec{\nabla} f(E,t) \right) \nonumber
\\   =I^{ph}\left\{f,n\right\}
\end{align}
where $E$ is the energy from the Fermi level, $\rho$ is the superconducting density of states (DoS) normalized by the normal-state DoS at the Fermi level, $D=v_F \ell/3$ is the normal-state diffusion coefficient, $I^{ph}$ the collision integral accounting for the interaction between quasiparticles and phonons, and $n$ is the phonon distribution function. Both $\rho$ the coherence factors $K^\pm$ can be obtained from the quasiclassical retarded Green's functions $g^R$ and $f^R$ as $\rho(E) = \Re g^R(E)$ and $K^\pm(E,E') = 1 \pm [\Re f^R(E) \Re f^R(E')]/[\Re g^R(E) \Re g^R(E')]$; 
approximate expressions for these Green's function will be discussed in Supplementary Note~\ref{app:optfluc}. As the presence of the Laplacian indicates, the functions appearing in Eq.~\eqref{eq: Diffusion equation} are in general dependent on the spatial coordinates. We will neglect this dependence by assuming diffusion to be sufficiently fast and averaging over the sample volume. In this context, fast diffusion means that the inverse of the diffusion time of a quasiparticle with energy $E>E_g$ over the spatial scale of the localized states is large compared to the localization rate; we consider this condition in more detail later in this Supplemental Note. 
The collision integral $I^{ph}=I^{ph}_e+I^{ph}_a+I^{ph}_g-I^{ph}_r$ can be separated into two scattering contributions (phonon emission $I^{ph}_e$ and absorption $I^{ph}_a$), a generation term $I^{ph}_g$, and a recombination one, $I^{ph}_r$. Introducing the short-hand notation
\begin{equation}\label{eq:Ipmdef}
    I_{\sigma'}^{\sigma}(E,\omega)=\omega^\mu\rho(\sigma'\omega-\sigma E)K^{\sigma}(E,\sigma'\omega-\sigma E)
\end{equation}
with $\sigma,\,\sigma'=\pm$ and $\mu=1,\,2,\,3$, these contributions are
\begin{widetext}
\begin{align}
    I^{ph}_e & = \frac{\rho(E)}{\tau_0 T_c^{\mu+1}} \int_0\!d\omega \, I_+^- (E,\omega) f(E+\omega)\left[1-f(E)\right]\left[n(\omega)+1\right] \label{eq:Ie} 
    -\frac{\rho(E)}{\tau_0 T_c^{\mu+1}}\int_0^E\!\!d\omega \, I_-^- (E,\omega) f(E)\left[1-f(E-\omega)\right]\left[n(\omega)+1\right] \\
    I^{ph}_a & = \frac{\rho(E)}{\tau_0 T_c^{\mu+1}} \int_0^E\!\!d\omega \, I_-^- (E,\omega) f(E-\omega)\left[1-f(E)\right]n(\omega) 
    - \frac{\rho(E)}{\tau_0 T_c^{\mu+1}} \int_0\!d\omega \, I_+^- (E,\omega) f(E)\left[1-f(E+\omega)\right]n(\omega) \label{eq:Iabs} \\
    I^{ph}_g & =  \frac{\rho(E)}{\tau_0 T_c^{\mu+1}} \int_E\!d\omega \, I_+^+ (E,\omega) 
    \left[1-f(E)\right]\!\left[1-f(\omega-E)\right]n(\omega) \label{eq:Igen} \\
    I^{ph}_r & = \frac{\rho(E)}{\tau_0 T_c^{\mu+1}} \int_E\!d\omega \, I_+^+ (E,\omega) 
    f(E) f(\omega-E)\left[n(\omega)+1\right]
\end{align}
\end{widetext}
where the time $\tau_0$ characterizes the strength of the electron-phonon interaction and the index $\mu$ parameterizes different regimes for such interaction: from a theoretical perspective, the case $\mu=2$ considered in the main text as well as in Refs.~\cite{Kaplan.1976} and \cite{Chang.1978} should apply in the ``clean'' case in which the typical phonon momentum $q$ is large on the scale of the inverse electron mean-free path for impurity scattering, $q \ell \gtrsim 1$. In the ``diffusive'' case $q \ell \ll 1$, the index should be $\mu=1$ for static impurities (not affected by the lattice movement due to phonons) and $\mu=3$ if the impurities move with the lattice~\cite{G.Catelani.2019,Giazotto.2006}.
In the main text we have assumed $\mu=2$; we present results for other values of $\mu$ in this Supplementary Note.

As discussed in the main text, we consider all states with energy below the Abrikosov-Gorkov gap $E_g$ to be localized. Consequently, the mobile and localized normalized densities are defined as
\begin{align}
    x_m & = \frac{2}{\Delta_0} \int_{E_g}\!dE \, \rho(E) f(E) \\
    x_l & = \frac{2}{\Delta_0} \int_0^{E_g}\!dE \, \rho(E) f(E)
\end{align}
Considering first the zero-temperature limit $n=0$, and assuming small quasiparticle occupation probability, $f \ll 1$, multiplying both sides of Eq.~\eqref{eq: Diffusion equation} by $2/\Delta_0$ and integrating over the relevant energy ranges we find
\begin{align}
    \frac{d x_m}{dt} & = -\Gamma_\mathrm{loc}x_m - r x_m x_l - r x_m^2 \\
    \frac{d x_l}{dt} & = \Gamma_\mathrm{loc}x_m - r x_m x_l - \Gamma_{ll} x_l^2
\end{align}
where, generalizing previous calculations (see for instance Ref.~\cite{Fischer.2023}) the recombination coefficient is approximately $r=\frac{2^\mu}{\tau_0}\left( \frac{\Delta}{T_c}\right)^{\mu+1}$. We note that the results $\Gamma_{mm}=\Gamma_{ml}=r$ is compatible with the arguments of Ref.~\cite{Bespalov.2016}, as one can see using their Eq.~(2) to calculate the recombination rate between a localized quasiparticle and the mobile ones (then the total recombination rate is found by multiplying that expression, where $p_1$ is inverse of the sample volume and $p_2 = p_{LO}$, by the number of mobile quasiparticles); neglecting any spatial dependence, we also obtain $\Gamma_{ll}=r$, but this equality may hold only for sufficiently large generation rate, see Ref.~\cite{Bespalov.2016} and the main text. It has been recently argued~\cite{deRooij.2024} (based in part on the results of Ref.~\cite{Kozorezov.2008} obtained in the limit of small impurity concentration) that the recombination rate between two quasiparticles trapped in the same trap could be faster than that between mobile quasiparticles. This does not affect our considerations: as we write in the main text, $\Gamma_{ll}$ effectively accounts for recombination between quasiparticles in different traps; in the ``bursting bubble'' model of Ref.~\cite{Bespalov.2016}, recombination between two overlapping quasiparticles takes place immediately, but the authors still find $\Gamma_{ll} < \Gamma_{mm}$ at low generation rate. It is in principle possible to consider general relationships between the three recombination coefficients; for our purposes, we note that any increase in the recombination coefficients involving localized quasiparticles can only lower the bound shown in Eq.~\eqref{eq:xl_ub} of the main text, which would imply more stringent conditions for localized-localized recombination to have an impact.

For the localization rate we have
\begin{equation}\label{eq:Gloc_supp}
    \Gamma_\mathrm{loc}x_m = \frac{2}{\Delta_0}\int\limits_{E_g}^\infty \!dE \, \rho(E)f(E)\tau^{-1}_\mathrm{loc}(E)
\end{equation}
with
\begin{equation}\label{eq:tauloc}
   \tau^{-1}_\mathrm{loc}(E) =  \frac{1}{\tau_0 T_c^{\mu+1}}\int_0^{E_g}\!dE' \, (E-E')^\mu \rho(E') K^{-}(E,E')
\end{equation}
An analogous expression was used in Ref.~\cite{Golubov.1993} to calculate the localization rate in a proximitized region with lower gap and in the core of a vortex.
Since the integration in Eq.~\eqref{eq:Gloc_supp} is over energy larger than $E_g$, we approximate the DoS with the Abirkosov-Gorkov $\rho_{AG}$, see Eq.~\eqref{eq:Glocdef} in the main text and the next Supplementary Note. Similarly, we use the exponentially decaying localized density of states $\rho_l$ [Eq.~\eqref{eq:rhol} and the next Supplementary Note] in Eq.~\eqref{eq:tauloc}. Then to evaluate the integral in its right-hand side, we note that under the assumption of low effective temperature $T_*$ of the mobile quasiparticles, $T_*/\Delta < \eta^{2/3}$, we can replace $E$ with $E_g$ in $K^-$; the same replacement can be used for $E'$ due to the exponential decay in $\rho_l$, and hence $K^- \simeq \eta^{2/3}$, see the next Supplementary Note and Ref.~\cite{Savich.2017}. With the above substitutions, the integral can be calculated explicitly and gives 
\begin{align}
    \tau^{-1}_\mathrm{loc}(\epsilon)=& \,r\frac{a_d 2^{2-\mu}}{8-d}\left(\frac{\epsilon_T}{\Delta} \right)^{3/2+\mu}\\ & \sum\limits_{k=0}^{\mu} \binom{\mu}{k} \epsilon^k\mathit{\Gamma}\left[ \frac{4\left( \alpha_d+ \mu + 1 -k\right)}{8-d}\right] \nonumber
\end{align}
with $\epsilon=(E-E_g)/\epsilon_T$, $\binom{\mu}{k}$ the binomial coefficient and $\mathit{\Gamma}$ the gamma function. For $\mu =2$ this expression reduces to Eq.~\eqref{eq:tloc_final} in the main text.

The relaxation rate $\tau_m^{-1}$ of a mobile quasiparticle of energy $E$ into a lower-energy mobile state can be obtained by changing the integration limits in the right-hand side of Eq.~\eqref{eq:tauloc} from $[0,E_g]$ to $[E_g,E]$; then, assuming $E/E_g -1 \ll \Delta\eta^{2/3}$ the expression to use for the density of states in the integrand is $\rho(E') \simeq \sqrt{2(E-E_g)/(3\Delta)}/\eta^{2/3}$, see e.g. Refs.~\cite{Larkin.1972,Savich.2017} and the next Supplementary Note. Performing the integral we obtain
\begin{equation}
    \tau_{m}^{-1}(E)=r \frac{2^{3/2-\mu}}{\sqrt{3}}\left(\frac{E-E_g}{E_g}\right)^{\mu+3/2} \sum\limits_{k=0}^{\mu} \binom{\mu}{k} \frac{(-1)^k}{(2k+3)}
\end{equation}

At finite temperature $T$, taking the phonon distribution function in the Bose-Einstein form $n(\omega) = (e^{\omega/T}-1)^{-1}$ and proceeding in a similar manner, from $I^{ph}_a$ in Eq.~\eqref{eq:Iabs} we find for the excitation rate when $\epsilon_T \ll T \ll \Delta\eta^{2/3}$
\begin{equation}
    \Gamma^T_\mathrm{ex} \simeq \frac{r}{2^\mu}\sqrt{\frac{2}{3}}\mathit{\Gamma}\left(\mu+\frac32\right)\zeta\left(\mu+\frac32\right) \left(\frac{T}{\Delta}\right)^{\mu+3/2}
\end{equation}
where $\zeta$ denotes the zeta function; the expression for $\Delta\eta^{2/3} \ll T \ll \Delta$ is found by multiplying the right-hand side by $\sqrt{3}/2$.

Another effect of finite temperature is to increase the localization rate due to stimulated phonon emission [the $n(\omega)$ terms in Eq.~\eqref{eq:Ie}], $\tau_\mathrm{loc}^{-1} \to \tau_\mathrm{loc}^{-1}  + \delta\tau_\mathrm{loc}^{-1} (T)$. For our purposes, it is sufficient to establish an upper bound on $\delta\tau_\mathrm{loc}^{-1}$, focusing on the regime $\epsilon_T \ll T \ll \Delta \eta^{2/3}$; the upper bound is obtained by noticing that $n(\omega) < T/\omega$ end hence
\begin{equation}
    \delta\tau^{-1}_{loc}(T)\Big|_{\mu} < \tau^{-1}_{loc}\Big|_{\mu-1}\frac{T}{\Delta}
\end{equation}
The bound is approximately saturated for $\epsilon \ll T/\epsilon_T$, which implies $\delta\tau_\mathrm{loc}^{-1} \gg \tau_\mathrm{loc}^{-1}$ for $1\ll \epsilon \ll T/\epsilon_T$; hence we can compare $\delta\tau_\mathrm{loc}$ to $\tau_m^{-1}$ to determine the crossover energy $\epsilon_c$ at finite temperature. Up to factors of order unity, we find $\epsilon_c \approx \left(T/\epsilon_T\right)^{2/5}$ independent of $\mu$. With this estimate, we can establish $\Gamma_\mathrm{loc} < r (\epsilon_T/\Delta)^{3/2+3\mu/5}(T/\Delta)^{2\mu/5} \ll \Gamma^{T}_\mathrm{ex}$.

Before moving on to considering the generation rates, let us comment here on the assumption of fast diffusion over the typical scale of the localized states. The latter have extensions of order the coherence length $\xi$, so we want to check when the condition $D(\epsilon)/\xi^2 \gg \tau_\mathrm{loc}^{-1}(\epsilon)$, with $D(\epsilon) \simeq D \rho(\epsilon) K^-(E_g,E_g)$ [cf. Eq.~\eqref{eq: Diffusion equation}]. 
Using the square-root threshold (AG) form for the density of states and $\xi \simeq \sqrt{D/\Delta}$, up to numerical factors of order one we can write the left-hand side of the condition as  $\Delta \sqrt{\epsilon_T\epsilon/\Delta}$. For the right-hand side, we consider as a worst case the finite-temperature upper bound discussed above, in which case the condition becomes $1 \gg (r/\Delta)^2 (\epsilon_T/\Delta)^{1+6\mu/5} (T/\Delta)^{4\mu/5-1}$, where we assumed that the typical quasiparticle energy is of order of temperature, $\epsilon \sim T/\epsilon_T$. 
For elemental, low $T_c$ superconductors, $r/\Delta$ ranges approximately between $10^{-3}$ for Al and $0.4$ for Nb~\cite{Kaplan.1976}; for $\mu>5/4$, all other terms on the right-hand side are also small, so the condition is satisfied. 
It is easy to check that even for $\mu=1$ the temperature $T$ is in practice always much larger than what would satisfy the inequality.

\subsection{Generation rates}

The generation rates $g_l$ and $g_m$ due to thermal phonons are obtained by integrating Eq.~\eqref{eq:Igen} over $E$ from $0$ to $E_g$ and from $E_g$ to infinity, respectively. We make the change of variables $\omega = E+E'$ and approximate $n(E+E') \simeq e^{-(E+E')/T}$, which is valid for $T\ll E_g$. We can also approximate $(E+E')^{\mu} \simeq (2E_g)^\mu$, and assume $T \ll \Delta \eta^{2/3}$ so that $K^+ \simeq 2$. After these approximations, the integrals factorize and we can write
\begin{equation}
    g_l = \mathrm{g}_l^2 + \mathrm{g}_l \mathrm{g}_m \, , \qquad g_m = \mathrm{g}_m^2 + \mathrm{g}_l \mathrm{g}_m
\end{equation}
with
\begin{equation}
    \mathrm{g}_l
    \simeq 2 \sqrt{r} \frac{1}{\Delta} \int_0^{E_g} dE \, \rho(E) e^{-E/T}
\end{equation}
while for $\mathrm{g}_m$ the integration is between $E_g$ and infinity. To evaluate the latter, we can take $\rho$ in the square-root threshold form to find
\begin{equation}
    \mathrm{g}_m \simeq \frac{1}{\eta^{2/3}} \sqrt{\frac{2\pi r}{3}}  \left(\frac{T}{\Delta}\right)^{3/2} e^{-E_g/T}
\end{equation}
For $\mathrm{g}_l$ we have
\begin{equation}
    \mathrm{g}_l \simeq \frac{2}{\eta^{2/3}} \sqrt{r} a_d \left(\frac{\epsilon_T}{\Delta}\right)^{3/2} e^{-E_g/T}
    \int_0  d\tilde{\epsilon} \, \tilde{\epsilon}^{\alpha_d} e^{-\tilde{\epsilon}^{2-d/4}} e^{\epsilon_T \tilde{\epsilon}/T}
\end{equation}
and we distinguish two cases. If $T \gg \epsilon_T$, we can approximate the last exponential factor by unity and the integral gives the numerical factor $4 \mathit{\Gamma}[4(\alpha_d+1)(8-d)]/(8-d)$; in this case $\mathrm{g}_l \ll \mathrm{g}_m$ and hence $g_l \ll g_m$. If $T\ll\epsilon_T$ the integral can be estimated by the Laplace method to find that the $\mathrm{g}_l$ is enhanced by an exponential factor with argument proportional to $(\epsilon_T/T)^{(8-d)/(4-d)}$; hence in this case $g_l \gg g_m$. For completeness, we recall the thermal equilibrium BCS result $g_m=r (2\pi T/\Delta) e^{-2\Delta/T}$, which approximately applies for $T \gtrsim \Delta\eta^{1/3}$; in this temperature regime, we have $g_m \gg g_l$.

Similar considerations can be made if considering generation by absorption of a photon of energy $\omega_0$ (by replacing $K^+$ with $K^-$ and $n(\omega)$ with a delta-function centered at $\omega_0$). It is possible to show that for $\omega_0 = 2E_g$ one finds $g_l = g_m$, while if $\omega_0 = 2E_g + \delta\omega$ with $\epsilon_T \ll \delta\omega \ll \Delta\eta^{2/3}$, then $g_l \ll g_m$. 

\subsection{Density decay rates}

Here we consider the decay rates of the quasiparticle densities for small deviations from the steady state. We consider only  the case $\kappa > \kappa_c$, so that we can set $\Gamma_{mm}=\Gamma_{ml}=\Gamma_{ll}=r$ in Eqs.~\eqref{eq: Effective Rate Equation} and \eqref{eq:xm rate eq} of the main text. Then writing $x_\alpha = \bar{x}_\alpha + \delta x_\alpha$, where $\bar{x}_\alpha$ are the steady-state solutions in Eq.~\eqref{eq:xmxlbetatilde} and $\delta x_\alpha \ll \bar{x}_\alpha$, keeping only terms linear in $\delta x_\alpha$ we find
\begin{align}
    \delta\dot{x}_l & = \Gamma_\mathrm{loc} \delta x_m -\Gamma_\mathrm{ex} \delta x_l - r \bar{x}_m \delta x_l - r \bar{x}_l \delta x_m - 2 r \bar{x}_l \delta x_l \\
    \delta\dot{x}_m \! & =  \Gamma_\mathrm{ex} \delta x_l - \Gamma_\mathrm{loc} \delta x_m - r \bar{x}_m \delta x_l - r \bar{x}_l \delta x_m - 2 r \bar{x}_m \delta x_m
\end{align}
From these equations it is straightforward to calculate the decay rates $\lambda_1=2r \bar{x}$ and $\lambda_2 = \Gamma_\mathrm{loc}+\Gamma_\mathrm{ex}+r\bar{x}$ given in the main text, which are a special case of the more general ones given in Ref.~\cite{Grunhaupt.2018}. One can check that $\lambda_1$ corresponds to the decay rate of the total density by taking the sum of the two equations above. Similarly, for $\lambda_2$ one can check the result by setting $\delta x_l = -\delta x_m$.

Our result shows that if $\Gamma_\mathrm{loc}+\Gamma_\mathrm{ex} \ll r \bar{x}$, the two modes have decay rates that differ only by a factor of 2, so they could be difficult to distinguish experimentally. Assuming that generation is dominated by thermal phonons and $T \gtrsim \Delta\eta^{1/3}$ (which also implies $\Gamma_\mathrm{ex} \gg \Gamma_\mathrm{loc}$), using Eq~\eqref{eq:Gex} we find $\Gamma_\mathrm{ex} \lesssim r\bar{x}$ for $T/\Delta \gtrsim 0.14$ (corresponding to $T \gtrsim 0.24 T_c$). Therefore, only in the temperature range $\epsilon_T \ll T \lesssim 0.24 T_c$ it is possible to discern the effect of excitation from localized state on the relaxation of the quasiparticle density; we note that the temperatures in Ref.~\cite{deRooij.2024} are clearly in the regime $T \gg \epsilon_T$ and extend roughly to the upper limit we estimate here, as the highest temperature reached there is $T\simeq 0.22~$K$\,\simeq 0.25 T_c$. We note, however, that in this regime the mobile density is much larger than the localized one, and in the experiment the resonator is driven resonantly at a frequency much larger than $\epsilon_T$, so it is unclear why the difference mode should be detected rather than the total one.

\section{Optimal fluctuations}
\label{app:optfluc}

The formula for the subgap DoS [Eq.~\eqref{eq:rhol} in the main text] is a long-established result in the literature, at least as far as the exponential factor is concerned~\cite{Larkin.1972,Silva.2005,Skvortsov.2013,Fominov.2016}. Here we revisit briefly its derivation to obtain the prefactor in front of the exponential for the two-dimensional case: the $d=3$ prefactor was calculated in Ref.~\cite{Larkin.1972}, and confirmed in Ref.~\cite{Silva.2005}, but the one for $d=2$ in the latter work is incorrect; we also include leading-order results for the coherence factors, which are used in Supplementary Note~\ref{app:genRT}.

The spatial fluctuations in the impurity concentration discussed in the main text can be parameterized by writing the pair-breaking parameter in the form $\eta(\vec{r}) = \eta +\delta\eta (\vec{r})$ with the fluctuating part being a random Gaussian variable~\cite{Silva.2005},
\begin{equation}
 \langle \delta\eta(\vec{r})\delta\eta(\vec{r}')\rangle = \frac{\eta^2}{n_\mathrm{imp}} \delta(\vec{r}-\vec{r}')
\end{equation}
with $n_\mathrm{imp}$ the average impurity concentration. The quasiclassical retarded Green's function can be determined by solving the so-called Usadel equation which, parameterizing the Green's function as~\cite{Skvortsov.2013} (we suppress the dependence on energy $E$ and position $\vec{r}$ for notational convenience)
\begin{equation}\label{eq: gf parametrization}
    g^R=i \sinh(\psi), \qquad
    f^R=i\cosh(\psi) 
\end{equation}
takes the form
\begin{equation}\label{eq: regular Part Usadel}
     -\xi^2\nabla_d^2 \psi + F(\psi)-\cosh(\psi)\sinh(\psi)\delta\eta(\vec{r}) = 0
\end{equation}
with $\xi=\sqrt{D/2\Delta}$ the coherence length, $\nabla^2_d$ the Laplacian in $d$ dimensions, and
\begin{equation}
    F(\psi)=\sinh(\psi)\left[1-\eta \cosh(\psi)\right]-\frac{E}{\Delta}\cosh(\psi)
\end{equation}
In principle, along with the Usadel equation one should also consider the self-consistent equation for the gap $\Delta$, which will display spatial fluctuations $\delta\Delta$ due to $\delta\eta(\vec{r})$; in fact, as shown in Ref.~\cite{Fominov.2016}, $\delta\Delta/\Delta \simeq \pi \delta\eta/(4-\pi\eta)$. In Eq.~\eqref{eq: regular Part Usadel}, such a fluctuation would contribute a term $(\delta\Delta/\Delta)\sinh(\psi)$ (we use here that $\Delta\eta =1/\tau_s$); in the energy region of interest (near $E_g$), this contribution is smaller by a factor $1/\cosh(\psi) \simeq \eta^{1/3}$ (see below) compared to the term proportional to $\delta\eta$ and can therefore be neglected for $\eta \ll 1$.

At leading order, we ignore the impurity fluctuations and denote the corresponding uniform (that is, Abrikosov-Gorkov) solution to $F(\psi)=0$ with $\psi_0$; at the gap edge, we have $\cosh \psi_0(E_g)= \eta^{-1/3}$. For energy close to $E_g$, the expansion of the functional $F$ for small $\delta\psi(E) = \psi(E)-\psi_0(E)$ takes the form~\cite{Skvortsov.2013} 
\begin{equation}
    F(\delta\psi) \simeq \Omega \delta\psi - \varrho \delta\psi^2
\end{equation}
where
\begin{equation}\label{eq:OmeRho}
\Omega = (1-\eta^{2/3})\sqrt{6\varepsilon}, \quad \varrho = \frac32\eta^{1/3}\sqrt{1-\eta^{2/3}}
\end{equation}
with $\varepsilon=1-E/E_g$. To proceed further, it is convenient to split $\delta\psi$ into its real and imaginary parts by defining
\begin{equation}
    \phi = \frac{\varrho}{\Omega}\Re \delta\psi \qquad \chi = \frac{\varrho}{\Omega}\Im\delta\psi
\end{equation}
and introduce the dimensionless coordinate $\tilde{\vec{r}} = \sqrt{\Omega} \, \vec{r}/\xi$. Then Eq.~\eqref{eq: regular Part Usadel} can be written as
\begin{align}
    \delta \eta & = \frac{\eta^{2/3}}{\sqrt{1-\eta^{2/3}}}\frac{\Omega^2}{\varrho}\left[-\tilde{\nabla}^2_d\phi + \tilde{F}(\phi)+\chi^2 \right] \label{eq:detaphi} \\
    0 & = \left[-\tilde{\nabla}^2_d + \tilde{F}'(\phi)\right] \chi \label{eq:usdelchi}
\end{align}
with $\tilde{F}(\phi)=\phi-\phi^2$

As discussed e.g. in Refs.~\cite{Silva.2005,Skvortsov.2013}, the exponential tail in the subgap DoS is determined by so-called optimal fluctuations, which are saddle points in the action $S_d = (n_\mathrm{imp}/2\eta^2)\int d^dr (\delta\eta)^2$. A non-trivial saddle point which is also a solution to the above equations can be obtained by setting $\chi=0$ and looking for a saddle point of $\tilde{S}_d = \frac{1}{2}\int d^d\tilde{r} [-\tilde{\nabla}_d^2\phi+\tilde{F}(\phi)]^2$. Such a saddle point, which we denote by $\bar{\phi}_d$, satisfies the equation
\begin{equation}\label{eq:saddlebarphi}
    \left[-\tilde{\nabla}^2_d+\tilde{F}'(\bar{\phi}_d)\right]\left[-\tilde{\nabla}^2_d\bar{\phi}_d+\tilde{F}(\bar{\phi}_d)\right] = 0
\end{equation}
Assuming rotational symmetry, in which case the Laplacian reduces to $\nabla_d^2 = r^{1-d}\partial_r r^{d-1}\partial_r$, this is equivalent to~\cite{Silva.2005,Skvortsov.2013}
\begin{equation}
    \left[-\tilde{\nabla}^2_{d+2}+\tilde{F}'(\bar{\phi}_d)\right]\left[-\tilde{\nabla}^2_{d-2}\bar{\phi}_d+\tilde{F}(\bar{\phi}_d)\right] = 0
\end{equation}
implying a sort of ``dimensional reduction'' by which the saddle point is solution to the non-linear differential equation in the second square brackets. For $d=3$, the resulting equation has the exact solution
\begin{equation}\label{eq:inst3D}
    \bar{\phi}_3 = \frac{3}{2 \cosh^2(\tilde{r}/2)}
\end{equation}
for which the action has the value $\tilde{S}_3=24\pi/5$.

\begin{figure}
    \centering
    \includegraphics[width=0.99\columnwidth]{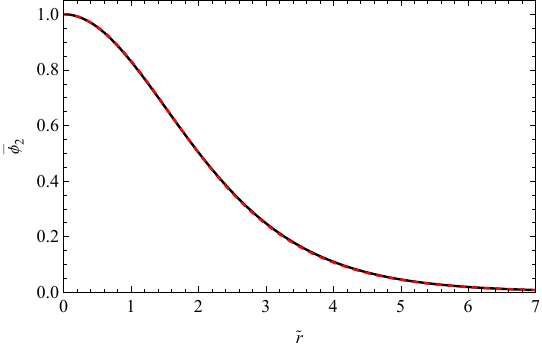}
    \caption{Solid line: numerical solution for $\bar\phi_2$. Dashed line: approximation using the variational Ansatz in Eq.~\eqref{eq:phi2var} with $a=0.2596$.}
    \label{fig:phi2}
\end{figure}

For $d=2$, we can find explicit expressions, up to unknown coefficients $a_1$ and $a_2$, in the limits $\tilde{r} \ll 1$ and $\tilde{r}\gg 1$. At small argument, we write $\bar{\phi}_2 = 1 - \delta\phi$, assume $\delta\phi \ll 1$, keep terms linear in $\delta\phi$ to find $\bar{\phi}_2 \simeq 1- a_1 \tilde{r} J_1(\tilde{r})\simeq 1- a_1 \tilde{r}^2/2$, with $J_1$ the Bessel function of the first kind. At large argument, we need $\bar{\phi}_2$ to decay quickly, so neglecting the term quadratic in $\bar{\phi}_2$ the approximate solution is $\bar{\phi}_2 \simeq a_2 \tilde{r} K_1(\tilde{r}) \propto \sqrt{\tilde{r}} e^{-\tilde{r}}$, with $K_1$ the modified Bessel function of the second kind. Together with Eq.~\eqref{eq:inst3D}, these approximate solutions suggest making the following variational Ansatz: 
\begin{equation}\label{eq:phi2var}
    \bar{\phi}_2 = \frac{(1+ a \tilde{r}^2)^{1/4}}{\cosh^2(\tilde{r}/2)}
\end{equation}
with $0<a<1$.
Numerically searching for a stationary point of the action, we find $a\simeq 0.2596$ and $\tilde{S}_2 \simeq 2.095$ (cf. Ref.~\cite{Skvortsov.2013}). The variational approximation is in good agreement with numerical solution to the saddle point equation, see Fig.~\ref{fig:phi2}. The numerical solution is found by discretization and using a shooting method in which the second derivative at the origin is used as a parameter to be chosen so that the solution at large $\tilde{r}$ decays to zero (the boundary conditions $\bar{\phi}_2(0)=1$ and $\partial_{\tilde{r}} \bar{\phi}_2(0)=0$ are kept fixed); in this way we find $\partial^2_{\tilde{r}} \bar{\phi}_2(0)\simeq -0.382$.
Using the expression in Eq.~\eqref{eq:phi2var}, it is in principle possible to adapt the arguments of Ref.~\cite{Bespalov.2016} to the $d=2$ case. In particular, one needs to calculate the function $g_2(\tilde{\vec{R}})= \int d^2 \,\tilde{r} \bar\phi_2(\tilde{\vec{r}}) \bar\phi_2(\tilde{\vec{r}}+\tilde{\vec{R}})$; its asymptotic behavior at large $\tilde{R}$ is $g_2(\tilde{R}) \simeq 2\pi \sqrt{\pi a} [\mathit{\Gamma}(7/4)/\mathit{\Gamma}(9/4)]\tilde{R}^{5/2}e^{-\tilde{R}}$. The latter expression shows that the overlap between trapped quasiparticles states decays more slowly in $d=2$ compared to the $d=3$ result $g_3(\tilde{R}) \propto e^{-\tilde{R}}$~\cite{Bespalov.2016}; therefore, the suppression of the recombination rate $\Gamma_{ll}$ below $\Gamma_{mm}$ as $\kappa$ (that is, the generation rate) decreases is weaker in $d=2$, implying that the recombination suppression is less efficient in thin superconducting films.
We note that despite this reduced efficiency, if the regime in which the considerations of Ref.~\cite{Bespalov.2016} apply were reached (implying $\kappa \ll 1$), the estimated density in films ($n_f$) would be higher than that in the bulk ($n_b$): the reference shows that in the bulk the density can be expressed as $n_b\sim 1/[r_c^3 \ln (1/\kappa)]$; the corresponding expression for thin films is $n_f \sim 1/[r_c^2 t \ln (t^2/\kappa r_c^2)]$, with thickness $t \ll \xi \lesssim r_c$. Using the latter condition, we find $n_b \ll n_f$.

For completeness, we also include here the $d=1$ case: considering small and large $\tilde{r}$, approximate solutions are of the form $\bar{\phi}_1 \simeq b_1 \left[1 - (1-b_1) \tilde{r}^2/2 + \ldots \right]$ and $\bar{\phi}_1 \propto \tilde{r} e^{-\tilde{r}}$. Then an appropriate variational Ansatz is
\begin{equation}\label{eq:ansatz1d}
    \bar{\phi}_1 = \frac{b\sqrt{1+(b-1/2)\tilde{r}^2}}{\cosh^2(\tilde{r}/2)}
\end{equation}
with $1/2 < b < 1$. The action has a stationary point at $b\simeq 0.695$ for which $\tilde{S}_1 \simeq 0.266$~\cite{Skvortsov.2013}, a result in good agreement with numerics from which we find $\bar{\phi}_1(0) \simeq 0.696$ [the agreement can be improved by using a two-parameter Ansatz in which the numerator on the right-hand side of Eq.~\eqref{eq:ansatz1d} is replaced by $b\sqrt{(1+(b+1/2)\tilde{r}^2+ c \tilde{r}^4)/(1+\tilde{r}^2)}$; the action then has a saddle point at $b\simeq 0.696$, $c\simeq 0.214$].

Using Eqs.~\eqref{eq:OmeRho} and \eqref{eq:detaphi}, we arrive at the following results for the action evaluated at the saddle point:
\begin{equation}
    S_d = \frac{16}{6^{d/4}\sigma^2}\left(1-\eta^{2/3}\right)^{2-d/2}\tilde{S}_d \, \varepsilon^{2-d/4}
\end{equation}
with $\sigma^2=\eta^{4/3}/(n_\mathrm{imp} \xi^d)$~\cite{Silva.2005,Skvortsov.2013}. We have substituted $\phi = \bar\phi_d$ and $\chi=0$ into  Eq.~\eqref{eq:detaphi}, thus finding the fluctuation $\delta\bar\eta$ in the pair-breaking parameter that gives the largest contribution to the action. Rewriting the expression above as $S_d = [(E_g-E)/\epsilon_T]^{2-d/4}$ we also obtain
\begin{equation}
    \epsilon_T = E_g \left[\frac{6^{d/4}\sigma^2}{16 (1-\eta^{2/3})^{2-d/2}\tilde{S}_d}\right]^{1/(2-d/4)}
\end{equation}
Note that through $\sigma$, $\epsilon_T$ depends not only on $\Delta$ and $\eta$, but also on $n_\mathrm{imp}$. One can show that the condition $n_\mathrm{imp} \xi^d \gg 1$ is sufficient to ensure that $\epsilon_T \ll \Delta - E_g$.

\subsection{Disorder averaging}

Having found the optimal fluctuation, we can now calculate the average over disorder of a quantity $y$, such as the density of states and the localization rate, written in terms of the Green's functions. The approach can be summarized as follows: the average over disorder is expressed as a functional integral over the (Gaussian) fluctuations $\delta\eta(\vec{r})$; a given disorder configuration can be approximated by dividing the systems in regions of volume $v$ containing a single bound state of energy $E_0 < E_g$, and approximating the fluctuation there with the ``best'' optimal fluctuation; the latter is found by minimizing the functional 
\begin{equation}
    D(\vec{r}'; \delta\eta)=\int d^d r \left[\delta\eta(\vec{r})-\delta\bar{\eta}(\vec{r}-\vec{r}')\right]^2
\end{equation}
with respect to $\vec{r}'$; we denote with $\vec{r}_0(\delta\eta(\vec{r}))$ the position of the minimum. Then the disorder-average of $y$ is given by
\begin{equation}
    \langle y \rangle= \Big\langle \int_v d^d r' \, y[\delta\eta(\vec{r})] \, \delta (\vec{r}'-\vec{r}_0(\delta\eta(\vec{r}))\Big\rangle
\end{equation}
where both $y$ and $\vec{r}_0$ depend on $\delta\eta$ (the former through the Green's functions)
Transforming the coordinates to $\nabla D(\vec{r}';\delta\eta)$ we have
\begin{equation}
    \langle y \rangle=\Big\langle \int_v d^d r'y \, \delta(\nabla D(\vec{r}';\delta\eta)) |\det \nabla\nabla D(\vec{r}';\delta\eta) |\Big\rangle
\end{equation}
Any fluctuation $\delta\eta$ can be expanded over an orthonormal basis $\varphi_j(\vec{r})$, $\delta\eta = \sum_{j=0} \zeta_j \varphi_j$, where we take $\varphi_0 = - a \delta\bar\eta(\vec{r}-\vec{r}')$ and  $\varphi_i = b\, \partial \delta\bar\eta(\vec{r}-\vec{r}')/\partial r_i$ for $i=1,\ldots, d$~\cite{Zittartz.1966}. The normalization constants are 
\begin{align}
 a^{-2} & = \int d^dr\, (\delta\bar\eta)^2 =   \frac{2\eta^{4/3}}{1-\eta^{2/3}} \frac{\Omega^{4-d/2} \xi^d}{\varrho^2} \tilde{S}_d \\
 b^{-2} & = \frac{1}{d} \int d^dr\, (\nabla\delta\bar\eta)^2 = \frac{2\eta^{4/3}}{1-\eta^{2/3}} \frac{\Omega^{5-d/2} \xi^{d-2}}{\varrho^2} \tilde{S}_d'
\end{align}
where
\begin{equation}
     \tilde{S}_d' = \frac{2}{d}\int d^d \tilde{r} \left[\frac{\partial}{\partial \tilde{r}} \left(\frac{\bar\phi_d'}{\tilde{r}}\right)\right]^2
\end{equation}
The corresponding values are $\tilde{S}_3' = 8\pi/7$, $\tilde{S}'_2 \simeq 0.408$, and $\tilde{S}_1' \simeq 0.022$.
From the above definitions, it follows that
\begin{equation}
    \nabla_i D = \frac{2\zeta_i}{b}, \quad |\det \nabla\nabla D | \simeq \left(\frac{2}{b^2}\right)^d
\end{equation}
and we have approximately
\begin{align}\label{eq:avy}
    \langle y \rangle & \simeq \frac{1}{b^d} \left(\frac{n_\mathrm{imp}}{2\pi \eta^2}\right)^{\frac{d+1}{2}} \\ 
    & \int_v d^d r' \int d\zeta_0 \, \Pi_i\!\int d\zeta_i \, y \, \delta(\zeta_i) e^{-\frac{n_\mathrm{imp}}{2\eta^2} (\zeta_0^2 + \sum_i \zeta_i^2)} \nonumber
\end{align}
where index $i$ runs from 1 to $d$ (that is, we assume that at leading order the contributions from higher-index terms give a factor of order unity; this is the case for the DoS and the quasiparticle transition rates).

The above results can be used to calculated the density of states; in this case for $E<E_g$ we have $y = \rho \simeq \cosh \psi_0\, \Im \delta \psi \simeq \frac{\Omega}{\eta^{1/3}\varrho} \chi$, and hence we need to determine how $\chi$ depends on $\delta\eta$. Looking at Eqs.~\eqref{eq:detaphi}-\eqref{eq:saddlebarphi} we expect $\chi \simeq \kappa \delta\bar\eta$; to find the proportionality coefficient $\kappa$, we use Eq.~\eqref{eq:detaphi} to express $\chi^2$ in terms of $\delta\eta$ and $\delta\bar\eta$ and write the result in the form $\chi^2 =\varrho\sqrt{1-\eta^{2/3}}/(\Omega^2\eta^{2/3})(\sum_j \zeta_j \varphi_j + \varphi_0/a)$. Multiplying both sides by $\phi_0$ and integrating over space we find $\kappa^2 = a^2 \varrho\sqrt{1-\eta^{2/3}}/(\Omega^2\eta^{2/3}) (\zeta_0+1/a)/\int d^dr \varphi_0^3 = -\varrho^4 (1-\eta^{2/3})^2/(a \eta^{8/3}\Omega^{8-d/2}\xi^d) (\zeta_0 + 1/a)/I_{3,d}$ with 
\begin{equation}
    I_{n,d} =-\int d^d \tilde{r} \left(\frac{2\bar\phi'}{\tilde{r}}\right)^n
\end{equation}
and we estimate 
\begin{equation}
I_{3,3} = 108\pi \left[ \frac{4\zeta(3)}{9\pi^2} + \frac{31\zeta(5)}{5\pi^4} - \frac{511\zeta(9)}{\pi^8}\right] \simeq 22.45
\end{equation}
(here $\zeta$ denotes Riemann's zeta function), $I_{3,2} \simeq 2.04$, and $I_{3,1} \simeq 0.18$. Using these results we have
\begin{align}
    \int d^dr \rho 
    & =\frac{\sqrt{1-\eta^{2/3}}\xi^{d/2}}{\sqrt{a} \, \eta \, \Omega^{1+d/4}} \frac{I_{1,d}}{\sqrt{I_{3,d}}}\sqrt{-\left(\zeta_0 + \frac1a\right)}
\end{align}
where $I_{1,3} = 24\pi$, $I_{2,1}\simeq 12.57$, and $I_{1,1} \simeq 1.86$.
Substitution of this expression into Eq.~\eqref{eq:avy} shows that we need to evaluate the integral
\begin{equation}
    \int_{-\infty}^{-1/a} \! d\zeta_0 \sqrt{-\left(\zeta_0 + \frac1a\right)} e^{-\frac{n_\mathrm{imp}}{2\eta^2} \zeta_0^2} \simeq \frac{\eta^3 a^{3/2}}{n_\mathrm{imp}^{3/2}} \frac{\sqrt{\pi}}{2} e^{-S_d}
\end{equation}
whose approximate value is calculated assuming $S_d \gg 1$, and therefore
\begin{align}
    \langle \rho \rangle & = \frac{a}{b^d} \frac{n_\mathrm{imp}^{(d-2)/2}\xi^{d/2}\sqrt{1-\eta^{2/3}}}{\pi^{d/2}2^{(d+3)/2}\eta^{d-1}\Omega^{1+d/4}} \frac{I_{1,d}}{\sqrt{I_{3,d}}} e^{-S_d} \\
    & = \frac{a_d}{\eta^{2/3}}\left(\frac{\epsilon_T}{E_g}\right)^{1/2}\tilde{\epsilon}^{\alpha_d}e^{-\tilde{\epsilon}^{2-d/4}}\left(1-\eta^{2/3}\right)^{3-7d/4+d^2/8}
\end{align}
with
\begin{equation}
a_d = \frac{\left(\tilde{S}'_d\right)^{d/2}I_{1,d}}{\sqrt{6}\pi^{d/2}\tilde{S}_d^{(d-1)/2}\sqrt{I_{3,d}}}    
\end{equation}
corresponding to $a_3 \simeq 0.53$, $a_2 \simeq 0.32$, and $a_1\simeq 0.15$. {We note that the coefficient for $d=3$ agrees with that in Ref.~\cite{Larkin.1972} (in Ref.~\cite{Bespalov.2016} an additional factor $2/3$ appears to be missing). Our result for $d=2$ differs from that in Ref.~\cite{Silva.2005}; in fact, the disagreement is not only in the numerical prefactor, but also in the parameter dependence: agreement in the latter can be recovered by correcting the factor $(2\pi\eta)^{-3/2}$ in Eq.~(B12) there, which should read $(2\pi\eta)^{-d/2}$. To our knowledge, the result for the exponential prefactor in $d=1$ is new.

Finally, let us comment on the calculation of the localization time $\tau_\mathrm{loc}$ in Eq.~\eqref{eq:tauloc}. To estimate the coherence factor we need the value of the ratio $\Re f^{R}/\Re g^{R}=\tanh\Re\psi$, which at leading order near $E=E_g$ equals $\sqrt{1-\eta^{2/3}}$; hence we conclude $K^- \simeq \eta^{2/3}$ (and $K^+ \simeq 2$).
For completeness, we note that for $E>E_g$ we have $\psi_0(E) = \psi_0(E_g) + \delta\psi_0$ with $\delta\psi_0 \simeq i \sqrt{2(E-E_g)/(3\Delta)}/[\eta^{1/3}(1-\eta^{2/3})^{1/4}]$, which gives rise to the square-root threshold behavior of the AG DoS. Since at leading order $K^-$ is independent of the fluctuations $\delta\eta$, the disorder-averaging acts only on $\Re g^{R}(E')$, leading to the density of states factor in Eq.~\eqref{eq:Ipmdef}. 


\end{document}